\documentclass[12pt]{article}
\usepackage[T1]{fontenc}
\usepackage[latin9]{inputenc}
\usepackage{color}
\usepackage{float}
\usepackage{url}
\usepackage{amsmath}
\usepackage{amssymb}
\usepackage{graphicx}
\usepackage{esint}
\usepackage[authoryear]{natbib}

\makeatletter

\providecommand{\tabularnewline}{\\}

\@ifundefined{date}{}{\date{}}

\usepackage{psfrag}\usepackage{epsf}\usepackage{enumerate}
\usepackage{multirow}
\usepackage{url}
\usepackage{mathrsfs}

\usepackage{latexsym}
\usepackage{times}
\usepackage{bm}
\usepackage{amsfonts}
\usepackage{multirow}
\usepackage{caption}
\usepackage{xr}

\bibpunct{(}{)}{,}{autheryear}{,}{,}

\newcommand{\biblist}{\begin{list}{}
{\listparindent 0.0cm \leftmargin 0.50cm \itemindent -0.50 cm
\labelwidth 0 cm \labelsep 0.50 cm
\usecounter{list}}\clubpenalty4000\widowpenalty4000}
\newcommand{\ebiblist}{\end{list}}

\usepackage{latexsym}
\usepackage{multirow}

\newtheorem{lemma}{Lemma}\newtheorem{theorem}{Theorem}\newtheorem{assumption}{Assumption}\newtheorem{remark}{Remark}\newtheorem{proposition}{Proposition}\newtheorem{example}{Example}\newcommand*{\indep}{%
	\rotatebox[origin=c]{90}{$\models$}
}

\newcommand{\pr}{P} 
\newcommand{\var}{\mathrm{var}}
\newcommand{\cov}{\mathrm{cov}}

\newcommand{\T}{\mathrm{\scriptscriptstyle T}}

\newcommand{\ipw}{\mathrm{IPW}}
\newcommand{\aipw}{\mathrm{AIPW}}
\newcommand{\ep}{\mathrm{ep}}

\newcommand{\reg}{\mathrm{reg}}
\newcommand{\mat}{\mathrm{mat}}
\newcommand{\opt}{\mathrm{opt}}

\newcommand{\J}{\mathcal{J}}
\newcommand{\N}{\mathcal{N}}
\newcommand{\plim}{\mathrm{plim}~}

\newcommand{\bone}{\mathbf{1}}

\newcommand{\blind}{0}

\makeatother

\begin{document}

\global\long\def\spacingset#1{
\global\long\def\baselinestretch{%
}%
\small\normalsize}%
 \spacingset{1}


\if0\blind { 
\title{\textbf{Combining multiple observational data sources to estimate
causal effects }}
\author{Shu Yang$^{*1}$ and Peng Ding$^{*2}$\\
 $^{*1}$Department of Statistics, North Carolina State University\\
 $^{*2}$Department of Statistics, University of California, Berkeley}

\maketitle
 } \fi

\if1\blind { \bigskip{}
\bigskip{}
\bigskip{}

\begin{center}
\textbf{\LARGE{}Combining multiple observational data sources to estimate
causal effects}\textbf{ }
\par\end{center}

\medskip{}
} \fi

\bigskip{}

\begin{abstract}
The era of big data has witnessed an increasing availability of multiple
data sources for statistical analyses. We consider estimation of causal
effects combining big main data with unmeasured confounders and smaller
validation data with supplementary information on these confounders.
Under the unconfoundedness assumption with completely observed confounders,
the smaller validation data allow for constructing consistent estimators
for causal effects, but the big main data can only give error-prone
estimators in general. However, by leveraging the information in the
big main data in a principled way, we can improve the estimation efficiencies
yet preserve the consistencies of the initial estimators based solely
on the validation data. Our framework applies to asymptotically normal
estimators, including the commonly-used regression imputation, weighting,
and matching estimators, and does not require a correct specification
of the model relating the unmeasured confounders to the observed variables.
We also propose appropriate bootstrap procedures, which makes our
method straightforward to implement using software routines for existing
estimators. 
\end{abstract}
\noindent \textit{Keywords:} Calibration; Causal inference; Inverse
probability weighting; Missing confounder; Two-phase sampling. \vfill{}

\newpage\spacingset{1.45} 

\section{Introduction}

Unmeasured confounding is an important and common problem in observational
studies. Many methods have been proposed to deal with unmeasured confounding
in causal inference, such as sensitivity analyses (e.g. \citealp{rosenbaum1983assessing}),
instrumental variable approaches (e.g. \citealp{angrist1996identification}),
etc. However, sensitivity analyses cannot provide point estimation,
and valid instrumental variables are often difficult to find in practice.
We consider the setting where external validation data provide additional
information on unmeasured confounders. To be more precise, the study
includes a large main dataset representing the population of interest
with unmeasured confounders and a smaller validation dataset with
additional information about these confounders.

Our framework covers two common types of studies. First, we have a
large main dataset, and then collect more information on unmeasured
confounders for a subset of units, e.g., using a two-phase sampling
design \citep{neyman1938contribution,cochran2007sampling,wang2009causal}.
Second, we have a smaller but carefully designed validation dataset
with rich covariates, and then link it to a larger main dataset with
fewer covariates. The second type of data is now ubiquitous. In the
era of big data, extremely large data have become available for research
purposes, such as electronic health records, claims databases, disease
data registries, census data, to name a few \citep[e.g.,][]{imbens1994combining,schneeweiss2005adjusting,chatterjee2016constrained}.
Although these datasets might not contain full confounder information
that guarantees consistent causal effect estimation, they can be useful
to increase efficiencies of statistical analyses.

In causal inference, \citet{sturmer2005adjusting} propose a propensity
score calibration method when the main data contain the outcome and
an error-prone propensity score adjusting for partial confounders,
and the validation data supplement a gold standard propensity score
adjusting for all confounders. \citet{sturmer2005adjusting} then
apply a regression calibration technique to correct for the measurement
error from the error-prone propensity score. This approach does not
require the validation data to contain the outcome variable. However,
this approach relies on the \textit{surrogacy property} entailing
that the outcome variable is conditionally independent of the error-prone
propensity score given the gold standard propensity score and treatment.
This surrogacy property is difficult to justify in practice, and its
violations can lead to substantial biases \citep{sturmer2007performance,lunt2012propensity}.
Under the Bayesian framework, \citet{mccandless2012adjustment} specify
a full parametric model of the joint distribution for the main and
validation data, and treat the gold standard propensity score as a
missing variable in the main data. \citet{antonelli2017guided} combines
ideas of Bayesian model averaging, confounder selection, and missing
data imputation into a single framework in this context. \citet{enders2018comparison}
use simulation to show that multiple imputation is more robust than
two-phase logistic regression against misspecification of imputation
models. \citet{lin2014adjustment} develop a two-stage calibration
method, which summarizes the confounding information through propensity
scores and combines the results from the main and validation data.
Their two-stage calibration focuses on the regression context with
a correctly specified outcome model. Unfortunately, regression parameters,
especially in the logistic regression model used by \citet{lin2014adjustment},
may not be the causal parameters of interest in general \citep{freedman2008randomization}.

In this article, we propose a general framework to estimate causal
effects in the setting where the big main data have unmeasured confounders,
but the smaller external validation data provide supplementary information
on these confounders. Under the assumption of ignorable treatment
assignment, causal effects can be identified and estimated from the
validation data, using commonly-used estimators, such as regression
imputation, (augmented) inverse probability weighting (\citealt{horvitz1952generalization,rosenbaum1983central};
\citealp{robins1994estimation,bang2005doubly,cao2009improving}),
and matching \citep[e.g.,][]{rubin1973matching,rosenbaum1989optimal,HeckmanIchimuraTodd1997,hirano2003efficient,hansen2004full,rubin2006matched,abadie2006large,stuart2010matching,abadie2016matching}.
However, these estimators based solely on the validation data may
not be efficient. We leverage the correlation between the initial
estimator from the validation data and the error-prone estimator from
the main data to improve the efficiency over the initial estimator.
This idea is similar to the two-stage calibration in \citet{lin2014adjustment};
however, their method focuses only on regression parameters and requires
the validation data to be a simple random sample from the main data.
Alternatively, the empirical likelihood is also an attractive approach
to combine multiple data sources (\citealp{chen1999pseudo,qin2000miscellanea,chen2002using}
and \citealp{chen2003information}). However, the empirical likelihood
approach needs sophisticated programming, and its computation can
be heavy when data become large. Our method is practically simple,
because we only need to compute commonly-used estimators that can
be easily implemented by existing software routines. Moreover, \citet{lin2014adjustment}
and the empirical likelihood approach can only deal with regular and
asymptotically linear (RAL) estimators often formulated by moment
conditions, but our framework can also deal with non-RAL estimators,
such as matching estimators. We also propose a unified bootstrap procedure
based on resampling the linear expansions of the estimators, which
is simple to implement and works for both RAL and matching estimators.

Furthermore, we relax the assumption that the validation data is a
random sample from the study population of interest. We also link
the proposed method to existing methods for missing data, viewing
the additional confounders as missing values for units outside of
the validation data. In contrast to most existing methods in the missing
data literature, the proposed method does not need to specify the
missing data model relating the unmeasured confounders with the observed
variables.

For simplicity of exposition, we use ``IID'' for ``identically
and independently distributed'', $1(\cdot)$ for the indicator function,
$\xi^{\otimes2}=\xi\xi^{\T}$ for a vector or matrix $\xi$, ``plim''
for the probability limit of a random sequence, and $A_{n}\cong B_{n}$
for two random sequences satisfying $A_{n}=B_{n}+o_{P}(n^{-1/2})$
with $n$ being the generic sample size. We relegate all regularity
conditions for asymptotic analyses to the on-line supplementary material.

\section{Basic setup\label{sec:Basic-setup}}

\subsection{Notation: causal effect and two data sources}

Following \citet{splawa1990application} and \citet{rubin1974estimating},
we use the potential outcomes framework to define causal effects.
Suppose that the treatment is a binary variable $A\in\{0,1\}$, with
$0$ and $1$ being the labels for control and active treatments,
respectively. For each level of treatment $a\in\{0,1\}$, we assume
that there exists a potential outcome $Y(a)$, representing the outcome
had the subject, possibly contrary to the fact, been given treatment
$a$. The observed outcome is $Y=Y(A)=AY(1)+(1-A)Y(0)$. Let a vector
of pretreatment covariates be $(X,U)$, where $X$ is observed for
all units, but $U$ may not be observed for some units.

Although we can extend our discussion to multiple data sources, for
simplicity of exposition, we first consider a study with two data
sources. The validation data have observations $\mathcal{O}_{2}=\{(A_{j},X_{j},U_{j},Y_{j}):j\in\mathcal{S}_{2}\}$
with sample size $n_{2}=|\mathcal{S}_{2}|$. The main data have observations
$\mathcal{O}_{1}=\{(A_{i},X_{i},Y_{i}):i\in\mathcal{S}_{1}\backslash\mathcal{S}_{2}\}\cup\mathcal{O}_{2}$
with sample size $n_{1}=|\mathcal{S}_{1}|$. In our formulation, we
consider the case with $\mathcal{S}_{2}\subset\mathcal{S}_{1}$, and
let $\rho=\lim_{n_{2}\rightarrow\infty}n_{2}/n_{1}\in[0,1]$. If one
has two separate main and validation datasets, the main dataset in
our context combines these two datasets. Although the main dataset
is larger, i.e., $n_{1}>n_{2}$, it does not contain full information
on important covariates $U$. Under a superpopulation model, we assume
that $\{A_{i},X_{i},U_{i},Y_{i}(0),Y_{i}(1):i\in\mathcal{S}_{1}\}$
are IID for all $i\in\mathcal{S}_{1}$, and therefore the observations
in $\mathcal{O}_{1}$ are also IID. The following assumption links
the main and validation data.

\begin{assumption}\label{assump: m}The index set $\mathcal{S}_{2}$
for the validation data of size $n_{2}$ is a simple random sample
from $\mathcal{S}_{1}$. \end{assumption}

Under Assumption \ref{assump: m}, $\{A_{j},X_{j},U_{j},Y_{j}(0),Y_{j}(1):j\in\mathcal{S}_{2}\}$
and the observations in $\mathcal{O}_{2}$ of the validation data
are also IID, respectively. We shall relax Assumption \ref{assump: m}
to allow $\mathcal{S}_{2}$ to be a general probability sample from
$\mathcal{S}_{1}$ in Section \ref{sec:Another-perspective}. But
Assumption \ref{assump: m} makes the presentation simpler.

\begin{example}\label{eg::twophase}

Two-phase sampling design is an example that results in the observed
data structure. In a study, some variables (e.g. $A,$ $X$, and $Y$)
may be relatively cheaper, while some variables (e.g. $U$) are more
expensive to obtain. A two-phase sampling design \citep{neyman1938contribution,cochran2007sampling,wang2009causal}
can reduce the cost of the study: in the first phase, the easy-to-obtain
variables are measured for all units, and in the second phase, additional
expensive variables are measured for a selected validation sample.
\end{example}

\begin{example}\label{eg::bigdata}Another example is highly relevant
in the era of big data, where one links small data with full information
on $(A,X,U,Y)$ to external big data with only $(A,X,Y)$. \citet{chatterjee2016constrained}
recently consider this scenario for parametric regression analyses.
\end{example}

Without loss of generality, we first consider the average causal effect
(ACE) 
\begin{equation}
\tau=E\{Y(1)-Y(0)\},\label{eq::ACE}
\end{equation}
and will discuss extensions to other causal estimands in Section \ref{subsec:Other-causal-estimands}.
Because of the IID assumption, we drop the indices $i$ and $j$ in
the expectations in \eqref{eq::ACE} and later equations.

In what follows, we define the conditional means of the outcome as
\[
\mu_{a}(X,U)=E(Y\mid A=a,X,U),\quad\mu_{a}(X)=E(Y\mid A=a,X),
\]
the conditional variances of the outcome as 
\[
\sigma_{a}^{2}(X,U)=\var(Y\mid A=a,X,U),\quad\sigma_{a}^{2}(X)=\var(Y\mid A=a,X),
\]
the conditional probabilities of the treatment as 
\[
e(X,U)=\pr(A=1\mid X,U),\quad e(X)=\pr(A=1\mid X).
\]

\subsection{Identification and model assumptions}

A fundamental problem in causal inference is that we can observe at
most one potential outcome for a unit. Following \citet{rosenbaum1983central},
we make the following assumptions to identify causal effects.

\begin{assumption}[Ignorability]\label{asump-ignorable} $Y(a)\indep A\mid(X,U)$
for $a=0$ and $1$.

\end{assumption}

Under Assumption \ref{asump-ignorable}, the treatment assignment
is ignorable in $\mathcal{O}_{2}$ given $(X,U)$. However, the treatment
assignment is only ``latent'' ignorable in $\mathcal{O}_{1}\backslash\mathcal{O}_{2}$
given $X$ and the latent variable $U$ \citep{frangakis1999addressing,jin2008principal}.

Moreover, we require adequate overlap between the treatment and control
covariate distributions, quantified by the following assumption on
the \textit{propensity score} $e(X,U)$.

\begin{assumption}[Overlap]\label{asump-overlap}There exist constants
$c_{1}$ and $c_{2}$ such that with probability $1$, $0<c_{1}\leq e(X,U)\leq c_{2}<1$.

\end{assumption}

Under Assumptions \ref{asump-ignorable} and \ref{asump-overlap},
$\pr\{A=1\mid X,U,Y(1)\}=\pr\{A=1\mid X,U,Y(0)\}=e(X,U)$, and $E\{Y(a)\mid X,U\}=E\{Y(a)\mid A=a,X,U\}=\mu_{a}(X,U)$.
The ACE $\tau$ can then be estimated through regression imputation,
inverse probability weighting (IPW), augmented inverse probability
weighting (AIPW), or matching. See \citet{rosenbaum2002observational},
\citet{imbens2004nonparametric} and \citet{rubin2006matched} for
surveys of these estimators.

In practice, the outcome distribution and the propensity score are
often unknown and therefore need to be modeled and estimated.

\begin{assumption}[Outcome model]\label{asump outcome}The parametric
model $\mu_{a}(X,U;\beta_{a})$ is a correct specification for $\mu_{a}(X,U)$,
for $a=0,1$; i.e., $\mu_{a}(X,U)=\mu_{a}(X,U;\beta_{a}^{*})$, where
$\beta_{a}^{*}$ is the true model parameter, for $a=0,1$.

\end{assumption}

\begin{assumption}[Propensity score model]\label{asump ps}The
parametric model $e(X,U;\alpha)$ is a correct specification for $e(X,U)$;
i.e., $e(X,U)=e(X,U;\alpha^{*})$, where $\alpha^{*}$ is the true
model parameter.

\end{assumption}

The consistency of different estimators requires different model assumptions.

\section{Methodology and important estimators\label{sec:Methodology}}

\subsection{Review of commonly-used estimators based on validation data\label{sec::estimators}}

The validation data $\{(A_{j},X_{j},U_{j},Y_{j}):j\in\mathcal{S}_{2}\}$
contain observations of all confounders $(X,U)$. Therefore, under
Assumptions \ref{asump-ignorable} and \ref{asump-overlap}, $\tau$
is identifiable and can be estimated by some commonly-used estimator
solely from the validation data, denoted by $\hat{\tau}_{2}$. Although
the main data do not contain the full confounding information, we
leverage the information on the common variables $(A,X,Y)$ as in
the main data to improve the efficiency of $\hat{\tau}_{2}$. Before
presenting the general theory, we first review important estimators
that are widely-used in practice.

Let $\mu_{a}(X,U;\beta_{a})$ be a working model for $\mu_{a}(X,U)$,
for $a=0,1$, and $e(X,U;\alpha)$ be a working model for $e(X,U)$.
We construct consistent estimators $\hat{\beta}_{a}$ $(a=0,1)$ and
$\hat{\alpha}$ based on $\mathcal{O}_{2}$, with probability limits
$\beta_{a}^{*}$ $(a=0,1)$ and $\alpha^{*}$, respectively. Under
Assumption \ref{asump outcome}, $\mu_{a}(X,U;\beta_{a}^{*})=\mu_{a}(X,U)$,
and under Assumption \ref{asump ps}, $e(X,U;\alpha^{*})=e(X,U)$.

\begin{example}[Regression imputation]\label{example reg}The regression
imputation estimator is $\hat{\tau}_{\reg,2}=n_{2}^{-1}\sum_{j\in\mathcal{S}_{2}}\hat{\tau}_{\reg,2,j},$
where 
\[
\hat{\tau}_{\reg,2,j}=\mu_{1}(X_{j},U_{j};\hat{\beta}_{1})-\mu_{0}(X_{j},U_{j};\hat{\beta}_{0}).
\]
$\hat{\tau}_{\reg,2}$ is consistent for $\tau$ under Assumption
\ref{asump outcome}.

\end{example}

\begin{example}[Inverse probability weighting]\label{example ipw}The
IPW estimator is $\hat{\tau}_{\ipw,2}=n_{2}^{-1}\sum_{j\in\mathcal{S}_{2}}\hat{\tau}_{\ipw,2,j}$,
where 
\[
\hat{\tau}_{\ipw,2,j}=\frac{A_{j}Y_{j}}{e(X_{j},U_{j};\hat{\alpha})}-\frac{(1-A_{j})Y_{j}}{1-e(X_{j},U_{j};\hat{\alpha})}.
\]
$\hat{\tau}_{\ipw,2}$ is consistent for $\tau$ under Assumption
\ref{asump ps}. \end{example}

The Horvitz--Thompson-type estimator $\hat{\tau}_{\ipw,2}$ has large
variability, and is often inferior to the Hajek-type estimator \citep{hajek1971comment}.
We do not present the Hajek-type estimator because we can improve
it by the AIPW estimator below. The AIPW estimator employs both the
propensity score and the outcome models.

\begin{example}[Augmented inverse probability weighting]\label{example aipw}
Define the residual outcome as $R_{j}=Y_{j}-\mu_{1}(X_{j},U_{j};\hat{\beta}_{1})$
for treated units and $R_{j}=Y_{j}-\mu_{0}(X_{j},U_{j};\hat{\beta}_{0})$
for control units. The AIPW estimator is $\hat{\tau}_{\aipw,2}=n_{2}^{-1}\sum_{j\in\mathcal{S}_{2}}\hat{\tau}_{\aipw,2,j}$,
where 
\begin{eqnarray}
\hat{\tau}_{\aipw,2,j} & = & \frac{A_{j}R_{j}}{e(X_{j},U_{j};\hat{\alpha})}+\mu_{1}(X_{j},U_{j};\hat{\beta}_{1})-\frac{(1-A_{j})R_{j}}{1-e(X_{j},U_{j};\hat{\alpha})}-\mu_{0}(X_{j},U_{j};\hat{\beta}_{0}).\label{eq::aipw-ind}
\end{eqnarray}
$\hat{\tau}_{\aipw,2}$ is doubly robust in the sense that it is consistent
if either Assumption \ref{asump outcome} or \ref{asump ps} holds.
Moreover, it is locally efficient if both Assumptions \ref{asump outcome}
and \ref{asump ps} hold \citep{bang2005doubly,tsiatis2007semiparametric,cao2009improving}.

\end{example}

Matching estimators are also widely used in practice. To fix ideas,
we consider matching with replacement with the number of matches fixed
at $M$. Matching estimators hinge on imputing the missing potential
outcome for each unit. To be precise, for unit $j$, the potential
outcome under $A_{j}$ is the observed outcome $Y_{j};$ the (counterfactual)
potential outcome under $1-A_{j}$ is not observed but can be imputed
by the average of the observed outcomes of the nearest $M$ units
with $1-A_{j}$. \textcolor{black}{Let these matched units for unit
$j$ be indexed by $\J_{d,V,j}$, where the subscripts $d$ and $V$
denote the dataset $\mathcal{O}_{d}$ and the matching variable $V$
(e.g. $V=(X,U)$), respectively. }Without loss of generality, we use
the Euclidean distance to determine neighbors; the discussion applies
to other distances \citep{abadie2006large}. Let $K_{d,V,j}=\sum_{l\in\mathcal{S}_{d}}1(j\in\J_{d,V,l})$
be the number of times that unit $j$ is used as a match based on
the matching variable $V$ in $\mathcal{O}_{d}$.

\begin{example}[Matching]\label{example: mat} Define the imputed
potential outcomes as 
\[
\hat{Y}_{j}(1)=\begin{cases}
M^{-1}\sum_{l\in\J_{2,(X,U),j}}Y_{l} & \text{if }A_{j}=0,\\
Y_{j} & \text{if }A_{j}=1,
\end{cases}\quad\hat{Y}_{j}(0)=\begin{cases}
Y_{j} & \text{if }A_{j}=0,\\
M^{-1}\sum_{l\in\J_{2,(X,U),j}}Y_{l} & \text{if }A_{j}=1.
\end{cases}
\]
Then the matching estimator of $\tau$ is 
\[
\hat{\tau}_{\mat,2}^{(0)}=n_{2}^{-1}\sum_{j\in\mathcal{S}_{2}}\{\hat{Y}_{j}(1)-\hat{Y}_{j}(0)\}=n_{2}^{-1}\sum_{j\in\mathcal{S}_{2}}(2A_{j}-1)\left(Y_{j}-M^{-1}\sum_{l\in\J_{2,(X,U),j}}Y_{l}\right).
\]
\citet{abadie2006large} obtain the decomposition: 
\[
n_{2}^{1/2}(\hat{\tau}_{\mat,2}^{(0)}-\tau)=B_{2}+D_{2},
\]
where 
\begin{eqnarray}
B_{2} & = & n_{2}^{-1/2}\sum_{j\in\mathcal{S}_{2}}(2A_{j}-1)\left[M^{-1}\sum_{l\in\J_{2,(X,U),j}}\left\{ \mu_{1-A_{j}}(X_{j},U_{j})-\mu_{1-A_{j}}(X_{l},U_{l})\right\} \right],\label{eq:B2}\\
D_{2} & = & n_{2}^{-1/2}\sum_{j\in\mathcal{S}_{2}}\left[\mu_{1}(X_{j},U_{j})-\mu_{0}(X_{j},U_{j})-\tau\right.\nonumber \\
 &  & +\left.(2A_{j}-1)\left\{ 1+M^{-1}K_{2,(X,U),j}\right\} \left\{ Y_{j}-\mu_{A_{j}}(X_{j},U_{j})\right\} \right].\nonumber 
\end{eqnarray}
The difference $\mu_{1-A_{j}}(X_{j},U_{j})-\mu_{1-A_{j}}(X_{l},U_{l})$
in (\ref{eq:B2}) accounts for the matching discrepancy, and therefore
$B_{2}$ contributes to the asymptotic bias of the matching estimator.
\citet{abadie2006large} show that the matching estimators are biased
with $p\geq2$. Let $\hat{\mu}_{a}(X,U)$ be an estimator for $\mu_{a}(X,U)$,
obtained either parametrically, e.g., by a linear regression estimator,
or nonparametrically, for $a=0,1$. \citet{abadie2006large} propose
a bias-corrected matching estimator 
\[
\hat{\tau}_{\mat,2}=\hat{\tau}_{\mat,2}^{(0)}-n_{2}^{-1/2}\hat{B}_{2},
\]
where $\hat{B}_{2}$ is an estimator for $B_{2}$ by replacing $\mu_{a}(X,U)$
with $\hat{\mu}_{a}(X,U)$.

\end{example}

\subsection{A general strategy\label{subsec:A-general-strategy}}

We give a general strategy for efficient estimation of the ACE by
utilizing both the main and validation data. In Sections \ref{subsec:RAL}
and \ref{subsec:Matching-estimators}, we will provide examples to
elucidate the proposed strategy with specific estimators.

Although the estimators based on the validation data $\mathcal{O}_{2}$
are consistent for $\tau$ under certain regularity conditions, they
are inefficient without using the main data $\mathcal{O}_{1}.$ However,
the main data $\mathcal{O}_{1}$ do not contain important confounders
$U$; if we naively use the estimators in Examples \ref{example reg}--\ref{example: mat}
with $U$ being empty, then the corresponding estimators can be inconsistent
for $\tau$ and thus are error-prone in general. Moreover, for robustness
consideration, we do not want to impose additional modeling assumptions
linking $U$ and $(A,X,Y)$.

Our strategy is straightforward: we apply the same error-prone procedure
to both the main and validation data. The key insight is that the
difference of the two error-prone estimates is consistent for $0$
and can be used to improve efficiency of the initial estimator due
to its association with $\hat{\tau}_{2}$. Let an error-prone estimator
of $\tau$ from the main data be $\hat{\tau}_{1,\ep}$, which converges
to some constant $\tau_{\ep}$, not necessarily the same as $\tau.$
Applying the same method to the validation data $\ensuremath{\{(A_{j},X_{j},Y_{j}):j\in\mathcal{S}_{2}\}}$,
we can obtain another error-prone estimator $\hat{\tau}_{2,\ep}$.
More generally, we can consider $\tau_{\ep}$ to be an $L$-dimensional
vector of parameters identifiable based on the joint distribution
of $(A,X,Y)$, and $\hat{\tau}_{1,\ep}$ and $\hat{\tau}_{2,\ep}$
to be the corresponding estimators from the main and validation data,
respectively. For example, $\hat{\tau}_{d,\text{ep}}$ can contain
estimators of $\tau$ using different methods based on $\mathcal{O}_{d}$.

We consider a class of estimators satisfying 
\begin{equation}
n_{2}^{1/2}\left(\begin{array}{c}
\hat{\tau}_{2}-\tau\\
\hat{\tau}_{2,\ep}-\hat{\tau}_{1,\ep}
\end{array}\right)\rightarrow\N\left\{ 0_{L+1},\left(\begin{array}{cc}
v_{2} & \Gamma^{\T}\\
\Gamma & V
\end{array}\right)\right\} ,\label{eq:asymp}
\end{equation}
in distribution, as $n_{2}\rightarrow\infty$, which is general enough
to include all the estimators reviewed in Examples \ref{example reg}--\ref{example: mat}.
Heuristically, if (\ref{eq:asymp}) holds exactly rather than asymptotically,
by the multivariate normal theory, we have the following the conditional
distribution 
\[
n_{2}^{1/2}(\hat{\tau}_{2}-\tau)\mid n_{2}^{1/2}(\hat{\tau}_{2,\ep}-\hat{\tau}_{1,\ep})\sim\mathcal{N}\left\{ n_{2}^{1/2}\Gamma^{\T}V^{-1}(\hat{\tau}_{2,\ep}-\hat{\tau}_{1,\ep}),v_{2}-\Gamma^{\T}V^{-1}\Gamma\right\} .
\]
Let $\hat{v}_{2},$ $\hat{\Gamma}$ and $\hat{V}$ be consistent estimators
for $v_{2},$ $\Gamma$ and $V$. We set $n_{2}^{1/2}(\hat{\tau}_{2}-\tau)$
to equal its estimated conditional mean $n_{2}^{1/2}\hat{\Gamma}^{\T}\hat{V}^{-1}(\hat{\tau}_{2,\ep}-\hat{\tau}_{1,\ep})$,
leading to an estimating equation for $\tau$: 
\[
n_{2}^{1/2}(\hat{\tau}_{2}-\tau)=n_{2}^{1/2}\hat{\Gamma}^{\T}\hat{V}^{-1}(\hat{\tau}_{2,\ep}-\hat{\tau}_{1,\ep}).
\]
Solving this equation for $\tau$, we obtain the estimator 
\begin{equation}
\hat{\tau}=\hat{\tau}_{2}-\hat{\Gamma}^{\T}\hat{V}^{-1}(\hat{\tau}_{2,\ep}-\hat{\tau}_{1,\ep}).\label{eq:proposed estimator}
\end{equation}

\begin{proposition}Under Assumption \ref{assump: m} and certain
regularity conditions, if (\ref{eq:asymp}) holds, then $\hat{\tau}$
is consistent for $\tau$, and 
\begin{equation}
n_{2}^{1/2}(\hat{\tau}-\tau)\rightarrow\N(0,v_{2}-\Gamma^{\T}V^{-1}\Gamma),\label{eq:asymp var}
\end{equation}
in distribution, as $n_{2}\rightarrow\infty$. Given a nonzero $\Gamma$,
the asymptotic variance, $v_{2}-\Gamma^{\T}V^{-1}\Gamma,$ is smaller
than the asymptotic variance of $\hat{\tau}_{2}$, $v_{2}$.

\end{proposition}

The consistency of $\hat{\tau}$ does not require any component in
$\hat{\tau}_{1,\ep}$ and $\hat{\tau}_{2,\ep}$ to correctly estimate
$\tau$. That is, these estimators can be error prone. The requirement
for the error-prone estimators is minimal, as long as they are consistent
for the same (finite) parameters. Under Assumption \ref{assump: m},
$\hat{\tau}_{1,\ep}-\hat{\tau}_{2,\ep}$ is consistent for a vector
of zeros, as $n_{2}\rightarrow\infty$.

We can estimate the asymptotic variance of $\hat{\tau}$ by 
\begin{equation}
\hat{v}=(\hat{v}_{2}-\hat{\Gamma}^{\T}\hat{V}^{-1}\hat{\Gamma})/n_{2}.\label{eq:VE}
\end{equation}

\begin{remark}\label{rmk:opt0}

We construct the error prone estimators $\hat{\tau}_{1,\ep}$ and
$\hat{\tau}_{2,\ep}$ based on $\mathcal{O}_{1}$ and $\mathcal{O}_{2}$,
respectively. Another intuitive way is to construct $\hat{\tau}_{1,\ep}$
and $\hat{\tau}_{2,\ep}$ based on $\mathcal{O}_{1}\backslash\mathcal{O}_{2}$
and $\mathcal{O}_{2}$, respectively. In general, we can construct
the error prone estimators based on different subsets of $\mathcal{O}_{1}$
and $\mathcal{O}_{2}$ as long as their difference converges in probability
to zero. 
We show in the supplementary material that our construction maximizes
the variance reduction for $\hat{\tau}_{2}$, $\Gamma^{\T}V^{-1}\Gamma$,
given the procedure of the error prone estimators.

\end{remark}

\begin{remark}\label{rmk:best} We can view (\ref{eq:proposed estimator})
as the best consistent estimator of $\tau$ among all linear combinations
$\{\hat{\tau}_{2}+\lambda^{\T}(\hat{\tau}_{2,\ep}-\hat{\tau}_{1,\ep}):\lambda\in\mathbb{R}^{L}\}$,
in the sense that (\ref{eq:proposed estimator}) achieves the minimal
asymptotic variance among this class of consistent estimators. Similar
ideas appeared in design-optimal regression estimation in survey sampling
\citep{deville1992calibration,fuller2009sampling}, regression analyses
(\citealp{chen2000unified}; \citealp{chen2002cox} and \citealp{wang2015semiparametric}),
improved prediction in high dimensional datasets \citep{boonstra2012incorporating},
and meta-analysis \citep{fibrinogen2009systematically}. In the supplementary
material, we show that the proposed estimator in \eqref{eq:proposed estimator}
is the best estimator of $\tau$ among the class of estimators $\{\hat{\tau}=f(\hat{\tau}_{2},\hat{\tau}_{1,\ep},\hat{\tau}_{2,\ep})$:
$f(x,y,z)$ is a smooth function of $(x,y,z)$, and $\hat{\tau}$
is consistent for $\tau\}$, in the sense that \eqref{eq:proposed estimator}
achieves the minimal asymptotic variance among this class. \end{remark}

\begin{remark}\label{choice of ep} The choice of the error-prone
estimators will affect the efficiency of $\hat{\tau}$. From (\ref{eq:asymp var}),
for a given $\hat{\tau}_{2}$, to improve the efficiency of $\hat{\tau}$
with a $1$-dimensional error-prone estimator, we would like this
estimator to have a small variance $V$ and a large correlation with
$\hat{\tau}_{2}$, $\Gamma$. In principle, increasing the dimension
of the error-prone estimator would not decrease the asymptotic efficiency
gain as shown in the supplementary material. However, it would also
increase the complexity of implementation and harm the finite sample
properties. To ``optimize'' the trade-off, we suggest choosing the
error-prone estimator to be the same type as the initial estimator
$\hat{\tau}_{2}$. For example, if $\hat{\tau}_{2}$ is an AIPW estimator,
we can choose $\hat{\tau}_{d,\ep}$ to be an AIPW estimator without
using $U$ in a possibly misspecified propensity score model. The
simulation in Section \ref{sec:Simulation} confirms that this choice
is reasonable. \end{remark}

To close this subsection, we comment on the existing literature and
the advantages of our strategy. The proposed estimator $\hat{\tau}$
in (\ref{eq:proposed estimator}) utilizes both the main and validation
data and improves the efficiency of the estimator based solely on
the validation data. In economics, \citet{imbens1994combining} propose
to use the generalized method of moments \citep{hansen1982large}
for utilizing the main data which provide moments of the marginal
distribution of some economic variables. In survey sampling, calibration
is a standard technique to integrate auxiliary information in estimation
or handle nonresponse; see, e.g., \citet{chen2000unified}, \citet{wu2001model},
\citet{kott2006using}, \citet{chang2008using} and \citet{kim2016calibrated}.
An important issue is how to specify optimal calibration equations;
see, for example, \citet{deville1992calibration,robins1994estimation,wu2001model},
and \citet{lumley2011connections}. Other researchers developed constrained
empirical likelihood methods to calibrate auxiliary information from
the main data; see, e.g., \citet{chen1999pseudo,qin2000miscellanea,chen2002using}
and \citet{chen2003information}.

Compared to these methods, the proposed framework is attractive because
it is simple to implement which requires only standard software routines
for existing methods, and it can deal with estimators that cannot
be derived from moment conditions, e.g., matching estimators. Moreover,
as some semiparametric methods, our framework does not require a correct
model specification of the relationship between unmeasured covariates
$U$ and measured variables $(A,X,Y)$.

\subsection{Regular asymptotically linear (RAL) estimators\label{subsec:RAL}}

We first elucidate the proposed method with RAL estimators.

From the validation data, we consider the case when $\hat{\tau}_{2}-\tau$
is RAL; i.e., it can be asymptotically approximated by a sum of IID
random vectors with mean $0$: 
\begin{equation}
\hat{\tau}_{2}-\tau\cong n_{2}^{-1}\sum_{j\in\mathcal{S}_{2}}\psi(A_{j},X_{j},U_{j},Y_{j}),\label{eq:RAL1}
\end{equation}
where $\{\psi(A_{j},X_{j},U_{j},Y_{j}):j\in\mathcal{S}_{2}\}$ are
IID with mean $0$. The random vector $\psi(A,X,U,Y)$ is called the
influence function of $\hat{\tau}_{2}$ with $E(\psi)=0$ and $E(\psi^{2})<\infty$
(e.g. \citealp{bickel1993efficient}). Regarding regularity conditions,
see, for example, \citet{newey1990semiparametric}.

Let $e(X;\gamma)$ be an error-prone propensity score model for $e(X)$,
and $\mu_{a}(X;\eta_{a})$ be an error-prone outcome regression model
for $\mu_{a}(X)$, for $a=0,1$. The corresponding error-prone estimators
of the ACE can be obtained from the main data $\mathcal{O}_{1}$ and
the validation data $\mathcal{O}_{2}$. We consider the case when
$\hat{\tau}_{d,\ep}$ is RAL: 
\begin{equation}
\hat{\tau}_{d,\ep}-\tau_{\ep}\cong n_{d}^{-1}\sum_{j\in\mathcal{S}_{d}}\phi(A_{j},X_{j},Y_{j}),\qquad(d=1,2),\label{eq:RAL2}
\end{equation}
where $\{\phi(A_{j},X_{j},Y_{j}):j\in\mathcal{S}_{d}\}$ are IID with
mean $0$.

\begin{theorem}\label{Thm for RAL}Under certain regularity conditions,
(\ref{eq:asymp}) holds for the RAL estimators (\ref{eq:RAL1}) and
(\ref{eq:RAL2}), where $v_{2}=\var\{\psi(A,X,U,Y)\},$ $\Gamma=(1-\rho)\cov\{\psi(A,X,U,Y),\phi(A,X,Y)\},$
and $V=(1-\rho)$ $\times\var\{\phi(A,X,Y)\}.$

\end{theorem}

To derive $\hat{\Gamma}$ and $\hat{V}$ for RAL estimators, let $\hat{\phi}_{d}(A,X,Y)$
and $\hat{\psi}(A,X,U,Y)$ be estimators of $\phi(A,X,Y)$ and $\psi(A,X,U,Y)$
by replacing $E(\cdot)$ with the empirical measure and unknown parameters
with their corresponding estimators. Note that the subscript $d$
in $\hat{\phi}_{d}(A,X,Y)$ indicates that it is obtained based on
$\mathcal{O}_{d}$. Then, we can estimate $\Gamma$ and $V$ by 
\begin{eqnarray*}
\hat{\Gamma} & = & \widehat{\cov}(\hat{\tau}_{2},\hat{\tau}_{2,\ep}-\hat{\tau}_{1,\ep})=\left(1-\frac{n_{2}}{n_{1}}\right)\frac{1}{n_{2}}\sum_{j\in S_{2}}\hat{\psi}(A_{j},X_{j},U_{j},Y_{j})\hat{\phi}_{2}(A_{j},X_{j},Y_{j}),\\
\hat{V} & = & \widehat{\var}(\hat{\tau}_{1,\ep}-\hat{\tau}_{2,\ep})=\left(1-\frac{n_{2}}{n_{1}}\right)\frac{1}{n_{1}}\sum_{i\in S_{1}}\left\{ \hat{\phi}_{1}(A_{i},X_{i},Y_{i})\right\} ^{\otimes2}.
\end{eqnarray*}
Finally, we can obtain the estimator and its variance estimator by
(\ref{eq:proposed estimator}) and (\ref{eq:VE}), respectively.

The commonly-used RAL estimators include the regression imputation
and (augmented) inverse probability weighting estimators. Because
the influence functions for $\hat{\tau}_{\reg,2}$ and $\hat{\tau}_{\ipw,2}$
are standard, we present the details in the supplementary material.
Below, we state only the influence function for $\hat{\tau}_{\aipw,2}$.

For the outcome model, let $S_{a}(A,X,U,Y;\beta_{a})$ be the estimating
function for $\beta_{a}$, e.g., 
\[
S_{a}(A,X,U,Y;\beta_{a})=\bone(A=a)\frac{\partial\mu_{a}(X,U;\beta_{a})}{\partial\beta_{a}}\{Y-\mu_{a}(X,U;\beta_{a})\},
\]
for $a=0,1$, which is a standard choice for the conditional mean
model. For the propensity score model, let $S(A,X,U;\alpha)$ be the
estimating function for $\alpha$, e.g., 
\[
S(A,X,U;\alpha)=\frac{A-e(X,U;\alpha)}{e(X,U;\alpha)\{1-e(X,U;\alpha)\}}\frac{\partial e(X,U;\alpha)}{\partial\alpha},
\]
which is the score function from the likelihood of a binary response
model.\textcolor{black}{{} Moreover, let $\Sigma_{\alpha\alpha}=E\left\{ \partial S(A,X,U;\alpha^{*})/\partial\alpha^{\T}\right\} $.}
In addition, let $\hat{\beta}_{a}$ $(a=0,1)$ and $\hat{\alpha}$
be the estimators solving the corresponding empirical estimating equations
based on $\mathcal{O}_{2}$, with probability limits $\beta_{a}^{*}$
$(a=0,1)$ and $\alpha^{*}$, respectively.

\begin{lemma}[Augmented inverse probability weighting]\label{lemma(AIPW)}
For simplicity, denote $e_{j}^{*}=e(X_{j},U_{j};\alpha^{*})$, $\dot{e}_{j}^{*}=\partial e(X_{j},U_{j};\alpha^{*})/\partial\alpha^{\T}$,
$S_{j}^{*}=S(A_{j},X_{j},U_{j};\alpha^{*})$, $\mu_{aj}^{*}=\mu_{a}(X_{j},U_{j};\beta_{a}^{*})$,
$\dot{\mu}_{aj}^{*}=\partial\mu_{a}(X_{j},U_{j};\beta_{a}^{*})/\partial\beta_{a}^{\T}$,
$S_{aj}^{*}=S_{a}(A_{j},X_{j},U_{j},Y_{j};\beta_{a}^{*})$, and $\dot{S}_{aj}^{*}=\partial S_{a}(A_{j},X_{j},U_{j},Y_{j};\beta_{a}^{*})/\partial\beta_{a}^{\T}$
for $a=0,1$. Under Assumption \ref{asump outcome} or \ref{asump ps},
$\hat{\tau}_{\aipw,2}$ has the influence function 
\begin{eqnarray}
\psi_{\aipw}(A_{j},X_{j},U_{j},Y_{j}) & = & \frac{A_{j}Y_{j}}{e_{j}^{*}}+\left(1-\frac{A_{j}}{e_{j}^{*}}\right)\mu_{1j}^{*}\nonumber \\
 &  & -\frac{(1-A_{j})Y_{j}}{1-e_{j}^{*}}-\left(1-\frac{1-A_{j}}{1-e_{j}^{*}}\right)\mu_{0j}^{*}-\tau-H_{\aipw}\Sigma_{\alpha\alpha}^{-1}S_{j}^{*}\nonumber \\
 &  & +E\left\{ \left(1-\frac{1-A}{1-e^{*}}\right)\dot{\mu}_{0}^{*}\right\} \left\{ E(\dot{S}_{0}^{*})\right\} ^{-1}S_{0j}^{*}\label{eq:aipw influence fctn0}\\
 &  & -E\left\{ \left(1-\frac{A}{e^{*}}\right)\dot{\mu}_{1}^{*}\right\} \left\{ E(\dot{S}_{1}^{*})\right\} ^{-1}S_{1j}^{*},\label{eq:aipw influence fctn}
\end{eqnarray}
where 
\[
H_{\aipw}=E\left[\left\{ \frac{A(Y-\mu_{1}^{*})}{(e^{*})^{2}}-\frac{(1-A)(Y-\mu_{0}^{*})}{(1-e^{*})^{2}}\right\} \dot{e}^{*}\right].
\]

\end{lemma}

Lemma \ref{lemma(AIPW)} follows from standard asymptotic theory,
but as far as we know it has not appeared in the literature. \citet{lunceford2004stratification}
suggest using the terms without (\ref{eq:aipw influence fctn0}) and
(\ref{eq:aipw influence fctn}) for $\psi_{\aipw}$, which, however,
works only when Assumption \ref{asump ps} holds. Otherwise, the resulting
variance estimator is not consistent in general, as shown by simulation
in \citet{funk2011doubly}. The correction terms in (\ref{eq:aipw influence fctn0})
and (\ref{eq:aipw influence fctn}) also make the variance estimator
doubly robust in the sense that the variance estimator for $\hat{\tau}_{\aipw,2}$
is consistent if either Assumption \ref{asump outcome} or \ref{asump ps}
holds, not necessarily both.

For error-prone estimators, we can obtain the influence functions
similarly. The subtlety is that both the propensity score and outcome
models can be misspecified. For simplicity of the presentation, we
defer the exact formulas to the online supplementary material.

\subsection{Matching estimators\label{subsec:Matching-estimators}}

We then elucidate the proposed method with non-RAL estimators. An
important class of non-RAL estimators for the ACE are the matching
estimators. The matching estimators are not regular estimators because
the functional forms are not smooth due to the fixed numbers of matches
\citep{abadie2008failure}. Continuing with Example \ref{example: mat},
\citet{abadie2006large} express the bias-corrected matching estimator
$\hat{\tau}_{\mat,2}$ in a linear form as 
\begin{equation}
\hat{\tau}_{\mat,2}-\tau\cong n_{2}^{-1}\sum_{j\in\mathcal{S}_{2}}\psi_{\mat,j},\label{eq:linear form1}
\end{equation}
where 
\begin{equation}
\psi_{\mat,j}=\mu_{1}(X_{j},U_{j})-\mu_{0}(X_{j},U_{j})-\tau+(2A_{j}-1)\left\{ 1+M^{-1}K_{2,(X,U),j}\right\} \left\{ Y_{j}-\mu_{A_{j}}(X_{j},U_{j})\right\} .\label{eq:psi_mat}
\end{equation}
Similarly, $\hat{\tau}_{\mat,d,\ep}$ has a linear form 
\begin{equation}
\hat{\tau}_{\mat,d,\ep}-\tau_{\ep}\cong n_{d}^{-1}\sum_{j\in\mathcal{S}_{d}}\phi_{\mat,d,j},\label{eq:linear form2}
\end{equation}
where 
\begin{equation}
\phi_{\mat,d,j}=\mu_{1}(X_{j})-\mu_{0}(X_{j})-\tau_{\ep}+(2A_{j}-1)\left(1+M^{-1}K_{d,X,j}\right)\left\{ Y_{j}-\mu_{A_{j}}(X_{j})\right\} .\label{eq:phi_mat}
\end{equation}

\begin{theorem}\label{Thm for mat} Under certain regularity conditions,
(\ref{eq:asymp}) holds for the matching estimators (\ref{eq:linear form1})
and (\ref{eq:linear form2}), where 
\begin{eqnarray*}
v_{2} & = & \var\left\{ \tau(X,U)\right\} +\plim\left[n_{2}^{-1}\sum_{j\in\mathcal{S}_{2}}\left\{ 1+M^{-1}K_{2,(X,U),j}\right\} ^{2}\sigma_{A_{j}}^{2}(X_{j},U_{j})\right],\\
\Gamma & = & (1-\rho)\left(\cov\left\{ \mu_{1}(X,U)-\mu_{0}(X,U),\mu_{1}(X)-\mu_{0}(X)\right\} \vphantom{\sum_{j}}\right.\\
 &  & +\left.\plim\left[n_{2}^{-1}\sum_{j\in\mathcal{S}_{2}}\left\{ 1+M^{-1}K_{2,(X,U),j}\right\} \left(1+M^{-1}K_{2,X,j}\right)\sigma_{A_{j}}^{2}(X_{j},U_{j})\right]\right),\\
V & = & \left(1-\rho\right)\left[\var\left\{ \mu_{1}(X)-\mu_{0}(X)\right\} +\plim\left\{ n_{2}^{-1}\sum_{j\in S_{2}}\left(1+M^{-1}K_{2,X,j}\right)^{2}\sigma_{A_{j}}^{2}(X_{j})\right\} \right].
\end{eqnarray*}
\end{theorem}

The existence of the probability limits in Theorem \ref{Thm for mat}
are guaranteed by the regularity conditions specified in the supplementary
material \citep[c.f.][]{abadie2006large}.

To estimate $(v_{2},\Gamma,V)$ in Theorem \ref{Thm for mat}, we
need to estimate the conditional mean and variance functions of the
outcome given covariates. Following \citet{abadie2006large}, we can
estimate these functions via matching units with the same treatment
level. We will discuss an alternative bootstrap strategy in the next
subsection.

\subsection{Bootstrap variance estimation\label{subsec:Bootstrap-variance-estimation}}

The asymptotic results in Theorems \ref{Thm for RAL} and \ref{Thm for mat}
allow for variance estimation of $\hat{\tau}$. In addition, we also
consider the bootstrap for variance estimation, which is simpler to
implement and often has better finite sample performances \citep{otsu2016bootstrap}.
This is particularly important for matching estimators because the
analytic variance formulas involve nonparametric estimation of the
conditional variances $\sigma_{a}^{2}(x,u)$ and $\sigma_{a}^{2}(x)$.

There are two approaches for obtaining bootstrap observations: (a)
the original observations; and (b) the asymptotic linear terms of
the proposed estimator. For RAL estimators, bootstrapping the original
observations will yield valid variance estimators \citep{efron1986bootstrap,shao2012jackknife}.
However, for matching estimators, \citet{abadie2008failure} show
that due to lack of smoothness in their functional form, the bootstrap
based on approach (a) does not apply for variance estimation. This
is mainly because the bootstrap based on approach (a) cannot preserve
the distribution of the numbers of times that the units are used as
matches. As a remedy, \citet{otsu2016bootstrap} propose to construct
the bootstrap counterparts by resampling based on approach (b) for
the matching estimator.

To unify the notation, let $\psi_{j}$ indicate $\psi(A_{j},X_{j},U_{j},Y_{j})$
for RAL $\hat{\tau}_{2}$ and $\psi_{\mat,j}$ for $\hat{\tau}_{\mat,2}$;
and similar definitions apply to $\phi_{d,j}$ $(d=1,2)$. Let $\hat{\psi}_{j}$
and $\hat{\phi}_{d,j}$ be their estimated version by replacing the
population quantities by the estimated quantities $(d=1,2)$. Following
\citet{otsu2016bootstrap}, for $b=1,\ldots,B$, we construct the
bootstrap replicates for the proposed estimators as follows: 
\begin{description}
\item [{Step$\ 1.$}] Sample $n_{1}$ units from $\mathcal{S}_{1}$ with
replacement as $\mathcal{S}_{1}^{*(b)}$, treat the units with observed
$U$ as the bootstrap validation data $\mathcal{S}_{2}^{*(b)}$. 
\item [{Step$\ 2.$}] Compute the bootstrap replicates of $\hat{\tau}_{2}-\tau$
and $\hat{\tau}_{d,\ep}-\tau_{\ep}$ as 
\begin{eqnarray*}
\hat{\tau}_{2}^{(b)}-\hat{\tau}_{2} & = & n_{2}^{-1}\sum_{j\in\mathcal{S}_{2}^{*(b)}}\hat{\psi}_{j},\\
\hat{\tau}_{d,\ep}^{(b)}-\hat{\tau}_{d,\ep} & = & n_{d}^{-1}\sum_{j\in\mathcal{S}_{d}^{*(b)}}\hat{\phi}_{d,j},\qquad(d=1,2).
\end{eqnarray*}
\end{description}
Based on the bootstrap replicates, we estimate $\Gamma,V$ and $v_{2}$
by 
\begin{eqnarray}
\hat{\Gamma} & = & (B-1)^{-1}\sum_{b=1}^{B}(\hat{\tau}_{2}^{(b)}-\hat{\tau}_{2})(\hat{\tau}_{2,\ep}^{(b)}-\hat{\tau}_{1,\ep}^{(b)}-\hat{\tau}_{2,\ep}+\hat{\tau}_{1,\ep}),\label{eq:gamma-hat}\\
\hat{V} & = & (B-1)^{-1}\sum_{b=1}^{B}(\hat{\tau}_{2,\ep}^{(b)}-\hat{\tau}_{1,\ep}^{(b)}-\hat{\tau}_{2,\ep}+\hat{\tau}_{1,\ep})^{\otimes2},\label{eq:V-hat}\\
\hat{v}_{2} & = & (B-1)^{-1}\sum_{b=1}^{B}(\hat{\tau}_{2}^{(b)}-\hat{\tau}_{2})^{2}.\label{eq:v2-hat}
\end{eqnarray}
Finally, we estimate the asymptotic variance of $\hat{\tau}$ by (\ref{eq:VE}),
i.e., $\hat{v}=(\hat{v}_{2}-\hat{\Gamma}^{\T}\hat{V}^{-1}\hat{\Gamma})/n_{2}.$

\begin{theorem}\label{Thm: boot}Under certain regularity conditions,
$(\hat{\Gamma},\hat{V},\hat{v}_{2},\hat{v})$ are consistent for $\{\Gamma,V,\var(\hat{\tau}_{2}),\var(\hat{\tau})\}$.

\end{theorem}

\begin{remark}

If the ratio of $n_{2}$ and $n_{1}$ is small, the above bootstrap
approach may be unstable, because it is likely that some bootstrap
validation data contain only a few or even zero observations. In this
case, we use an alternative bootstrap approach, where we sample $n_{2}$
units from $\mathcal{S}_{2}$ with replacement as $\mathcal{S}_{2}^{*}$,
sample $n_{1}-n_{2}$ units from $\mathcal{S}_{1}\backslash\mathcal{S}_{2}$
with replacement, combined with $\mathcal{S}_{2}^{*}$, as $\mathcal{S}_{1}^{*}$,
and obtain the proposed estimators based on $\mathcal{S}_{1}^{*}$
and $\mathcal{S}_{2}^{*}$. This approach grantees that the bootstrap
validation data contain $n_{2}$ observations.

\end{remark}

\begin{remark}It is worthwhile to comment on a computational issue.
When the main data have a substantially large size, the computation
for the bootstrap can be demanding if we follow Steps 1 and 2 above.
In this case, we can use subsampling \citep{politis1999subsampling}
or the Bag of Little Bootstrap \citep{kleiner2014scalable} to reduce
the computational burden. More interestingly, when $n_{2}\rightarrow\infty$
and $\rho=0$, i.e., the validation data contain a small fraction
of the main data, $\Gamma$ and $V$ reduce to $\cov(\hat{\tau}_{2},\hat{\tau}_{2,\ep})$
and $\var(\hat{\tau}_{2,\ep})$, respectively. That is, when the size
of the main data is substantially large, we can ignore the uncertainty
of $\hat{\tau}_{1,\ep}$ and treat it as a constant, which is a regime
recently considered by \citet{chatterjee2016constrained}. In this
case, we need only to bootstrap the validation data, which is computationally
simpler. \end{remark}

\section{Extensions\label{sec:Extension}}

\subsection{Other causal estimands\label{subsec:Other-causal-estimands}}

Our strategy extends to a wide class of causal estimands, as long
as (\ref{eq:asymp}) holds. For example, we can consider the average
causal effects over a subset of population \citep{crump2006moving,li2016balancing},
including the average causal effect on the treated.

We can also consider nonlinear causal estimands. For example, for
a binary outcome, the log of the causal risk ratio is 
\[
\log\text{CRR}=\log\frac{\pr\{Y(1)=1\}}{\pr\{Y(0)=1\}}=\log\frac{E\{Y(1)\}}{E\{Y(0)\}},
\]
and the log of the causal odds ratio is 
\[
\log\text{COR}=\log\frac{\pr\{Y(1)=1\}/\pr\{Y(1)=0\}}{\pr\{Y(0)=1\}/\pr\{Y(0)=0\}}=\log\frac{E\{Y(1)\}/[1-E\{Y(1)\}]}{E\{Y(0)\}/[1-E\{Y(0)\}]}.
\]
We give a brief discussion for the $\log\text{CRR}$ as an illustration.
The key insight is that under Assumptions \ref{asump-ignorable} and
\ref{asump-overlap}, we can estimate $E\{Y(a)\}$ with commonly-used
estimators from $\mathcal{O}_{2}$, denoted by $\hat{E}\{Y(a)\}$,
for $a=0,1$. We can then obtain an estimator for the $\log\text{CRR}$
as $\log[\hat{E}\{Y(1)\}/\hat{E}\{Y(0)\}]$. Similarly, we can obtain
error-prone estimators for the $\log\text{CRR}$ from both $\mathcal{O}_{1}$
and $\mathcal{O}_{2}$ using only covariates $X$. By the Taylor expansion,
we can linearize these estimators and establish a similar result as
(\ref{eq:asymp}), which serves as the basis to construct an improved
estimator for the $\log\text{CRR}$.

\subsection{Design issue: optimal sample size allocation }

As a design issue, we consider planning a study to obtain the data
structure in Section \ref{sec:Basic-setup} subject to a cost constraint.
The goal is to find the optimal design, specifically the sample allocation,
that minimizes the variance of the proposed estimator subject to a
cost constraint, as in the classical two-phase sampling \citep{cochran2007sampling}.

Suppose that it costs $C_{1}$ to collect $(A,X,Y)$ for each unit,
and $C_{2}$ to collect $U$ for each unit. Thus, the total cost of
the study is 
\begin{equation}
C=n_{1}C_{1}+n_{2}C_{2}.\label{eq:budget}
\end{equation}
The variance of the proposed estimator $\hat{\tau}$ is of the form
\begin{equation}
n_{2}^{-1}v_{2}-(n_{2}^{-1}-n_{1}^{-1})\gamma,\label{eq:var}
\end{equation}
e.g., for RAL estimators, 
\begin{eqnarray*}
\gamma & = & \cov\{\psi(A,X,U,Y),\phi(A,X,Y)\}^{\T}\left[\var\{\phi(A,X,Y)\}\right]^{-1}\cov\{\psi(A,X,U,Y),\phi(A,X,Y)\}
\end{eqnarray*}
is the variance of the projection of $\psi(A,X,U,Y)$ onto the linear
space spanned by $\phi(A,X,Y).$ Minimizing (\ref{eq:var}) with respect
to $n_{1}$ and $n_{2}$ subject to the constraint (\ref{eq:budget})
yields the optimal $n_{1}^{*}$ and $n_{2}^{*}$, which satisfy 
\begin{equation}
\rho^{*}=\frac{n_{2}^{*}}{n_{1}^{*}}=\left\{ (1-R_{\psi|\phi}^{2})\times\frac{C_{1}}{C_{2}}\right\} ^{1/2},\label{eq:optimal sample alloc}
\end{equation}
where $R_{\psi|\phi}^{2}=\gamma/v_{2}$ is the squared multiple correlation
coefficient of $\psi(A,X,U,Y)$ on $\phi(A,X,Y)$, which measures
the association between the initial estimator and the error-prone
estimator. We derive \eqref{eq:optimal sample alloc} by the Lagrange
multipliers, and relegate the details to the supplementary material.
Not surprisingly, \eqref{eq:optimal sample alloc} shows that the
sizes of the validation data and the main data should be inversely
proportional to the square-root of the costs. In addition, from \eqref{eq:optimal sample alloc},
a large size $n_{2}$ for the validation data is more desirable when
the association between the initial estimator and the error-prone
estimator is small.

\subsection{Multiple data sources}

\textcolor{black}{We have considered the setting with two data sources,
and we can easily extend the theory to the setting with multiple data
sources $\mathcal{O}_{1},...,\mathcal{O}_{K}$, where $\mathcal{O}_{1},...,\mathcal{O}_{K-1}$
contain partial covariate information, and the validation data, $\mathcal{O}_{K}$,
contain full information for $(A,X,U,Y)$. For example, for $d=1,\ldots,K-1$,
$\mathcal{O}_{d}$ contains variables $(A,V_{d},Y)$ where $V_{d}\subsetneqq(X,U)$.
}Each dataset $\mathcal{O}_{d}$, indexed by $\mathcal{S}_{d}$, has
size $n_{d}$ for $d=1,\ldots,K$. This type of data structure arises
from a multi-phase sampling as an extension of Example \ref{eg::twophase}
or multiple sources of ``big data'' as an extension of Example \ref{eg::bigdata}.

Let $\hat{\tau}_{K}$ be the initial estimator for $\tau$ from the
validation data $\mathcal{O}_{K}$, and $\hat{\tau}_{d,\ep}$ be the
error-prone estimator for $\tau$ from $\mathcal{O}_{d}$ ($d=1,\ldots,K-1$).
Let $\hat{\tau}_{d,K,\ep}$ be the estimator obtained by applying
the same error-prone estimator for $O_{d}$ to $\mathcal{O}_{K}$,
so that $\hat{\tau}_{d,\ep}-\hat{\tau}_{d,K,\ep}$ is consistent for
$0$, for $d=1,\ldots,K-1$. Assume that 
\[
n_{K}^{1/2}\left(\begin{array}{c}
\hat{\tau}_{K}-\tau\\
\hat{\tau}_{1,\ep}-\hat{\tau}_{1,K,\ep}\\
\vdots\\
\hat{\tau}_{K-1,\ep}-\hat{\tau}_{K-1,K,\ep}
\end{array}\right)\rightarrow\N\left\{ \left(\begin{array}{c}
0\\
0_{L}
\end{array}\right),\left(\begin{array}{cc}
v_{K} & \Gamma^{\T}\\
\Gamma & V
\end{array}\right)\right\} ,
\]
in distribution, as $n_{K}\rightarrow\infty$, where $L=\sum_{d=1}^{K-1}\dim(\hat{\tau}_{d,\ep})$.
If $\Gamma$ and $V$ have consistent estimators $\hat{\Gamma}$ and
$\hat{V}$, respectively, then, extending the proposed method in Section
\ref{sec:Methodology}, we can use 
\[
\hat{\tau}=\hat{\tau}_{K}-\hat{\Gamma}^{\T}\hat{V}^{-1}\left(\begin{array}{ccc}
\hat{\tau}_{1,\ep}^{\T}-\hat{\tau}_{1,K,\ep}^{\T}, & \cdots, & \hat{\tau}_{K-1,\ep}^{\T}-\hat{\tau}_{K-1,K,\ep}^{\T}\end{array}\right)^{\T}
\]
to estimate $\tau$. The estimator $\hat{\tau}$ is consistent for
$\tau$ with the asymptotic variance $v_{K}-\Gamma^{\T}V^{-1}\Gamma$,
which is smaller than the asymptotic variance of $\hat{\tau}_{K}$,
$v_{K}$, if $\Gamma$ is non-zero. Similar to the reasoning in Remark
\ref{choice of ep}, using more data sources will improve the asymptotic
estimation efficiency of $\tau.$

\section{Simulation\label{sec:Simulation}}

In this section, we conduct a simulation study to evaluate the finite
sample performance of the proposed estimators. In our data generating
model, the covariates are $X_{i}\sim$Unif$(0,2)$ and $U_{i}=0.5+0.5X_{i}-2\sin(X_{i})+2\mathrm{sign}\{\sin(5X_{i})\}+\epsilon_{i}$,
where $\epsilon_{i}\sim$Unif$(-0.5,0.5)$. The potential outcomes
are $Y_{i}(0)=-X_{i}-U_{i}+\epsilon_{i}(0)$ and $Y_{i}(1)=-X_{i}+4U_{i}+\epsilon_{i}(1)$,
where $\epsilon_{i}(0)\sim\N(0,1)$, $\epsilon_{i}(1)\sim\N(0,1)$,
and $\epsilon_{i}(0)$ and $\epsilon_{i}(1)$ are independent. Therefore,
the true value of the ACE is $\tau=E(5U_{i})$. The treatment indicator
$A_{i}$ follows Bernoulli($\pi_{i})$ with logit$(\pi_{i})=1-0.5X_{i}-0.5U_{i}$.
The main data $\mathcal{O}_{1}$ consist of $n_{1}$ units, and the
validation data $\mathcal{O}_{2}$ consist of $n_{2}$ units randomly
selected from the main data.

The initial estimators are the regression imputation, (A)IPW and matching
estimators applied solely to the validation data, denoted by $\hat{\tau}_{\reg,2}$,
$\hat{\tau}_{\ipw,2}$, $\hat{\tau}_{\aipw,2}$ and $\hat{\tau}_{\mat,2}$,
respectively. To distinguish the estimators constructed based on different
error-prone methods, we assign each proposed estimator a name with
the form $\hat{\tau}_{\mathrm{method},2\text{\&}\mathrm{methods}}$,
where ``method,2'' indicates the initial estimator applied to the
validation data $\mathcal{O}_{2}$, and ``methods'' indicates the
error-prone estimator(s) used to improve the efficiency of the initial
estimator. For example, $\hat{\tau}_{\reg,2\&\ipw}$ indicates the
initial estimator is the regression imputation estimator and the error-prone
estimator is the IPW estimator. We compare the proposed estimators
with the initial estimators in terms of percentages of reduction of
mean squared errors, defined as $\{1-\text{MSE}(\hat{\tau}_{\mathrm{method,}2\text{\&}\mathrm{methods}})/\text{MSE}(\hat{\tau}_{\mathrm{method},2})\}\times100\%$.
To demonstrate the robustness of the proposed estimator against misspecification
of the imputation model, we consider the multiple imputation (MI,
\citealp{rubin1987multiple}) estimator, denoted by $\hat{\tau}_{\mathrm{mi}}$,
which uses a regression model of $U$ given $(A,X,Y)$ for imputation.
We implement MI using the ``mice'' package in R with $m=10$.

Based on a point estimate $\hat{\tau}$ and a variance estimate $\hat{v}$
obtained by the asymptotic variance formula or the bootstrap method
described in Section \ref{subsec:Bootstrap-variance-estimation},
we construct a Wald-type $95\%$ confidence interval $(\hat{\tau}-z_{0.975}\hat{v}^{1/2},\hat{\tau}+z_{0.975}\hat{v}^{1/2})$,
where $z_{0.975}$ is the $97.5\%$ quantile of the standard normal
distribution. We further compare the variance estimators in terms
of empirical coverage rates.

\begin{figure}
\begin{centering}
\includegraphics[scale=0.7]{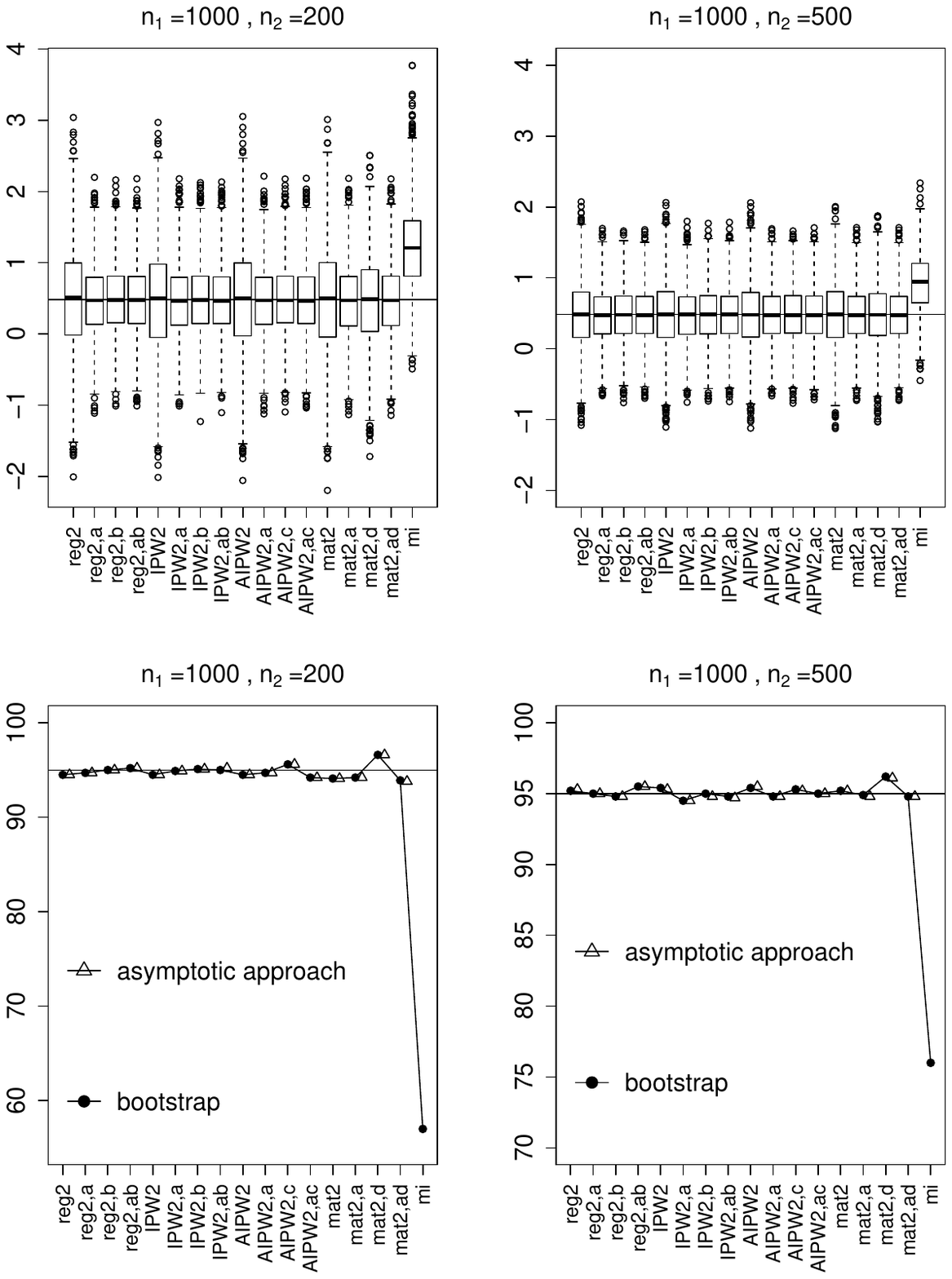} 
\par\end{centering}
 \caption{\label{fig:1}Simulation results of point estimates (top panels) and
coverage rates (bottom panels): the subscripts "a", "b", "c",
"d" stand for methods "reg", "IPW", "AIPW", "mat", respectively,
"reg2" is $\hat{\tau}_{\reg,2}$, "reg2,method" is $\hat{\tau}_{\reg,2\&\text{method}}$,
other notation is defined similarly, and "mi" is $\hat{\tau}_{\mathrm{mi}}$.}
\end{figure}

Figure \ref{fig:1} shows the simulation results over $2,000$ Monte
Carlo samples for $(n_{1},n_{2})=(1000,$ $200)$ and $(n_{1},n_{2})=(1000,500)$.
The multiple imputation estimator is biased due to the missipsecification
of the imputation model. In all scenarios, the proposed estimators
are unbiased and improve the initial estimators. Using the error-prone
estimator of the same type of the initial estimator achieves a substantial
efficiency gain, and the efficiency gain from incorporating additional
error-prone estimator is not significantly important. Because of the
practical simplicity, we recommend using the same type of error-prone
estimator to improve the efficiency of the initial estimator. 
Confidence intervals constructed from the asymptotic variance formula
and the bootstrap method work well, in the sense that the empirical
coverage rate of the confidence intervals is close to the nominal
coverage rate. In our settings, the matching estimator has the smallest
efficiency gain among all types of estimators.

\section{Application\label{sec:Real-data-application}}

We present an analysis to evaluate the effect of chronic obstructive
pulmonary disease (COPD) on the development of herpes zoster (HZ).
COPD is a chronic inflammatory lung disease that causes obstructed
airflow from the lungs, which can cause systematic inflammation and
dysregulate a patient's immune function. The hypothesis is that people
with COPD are at increased risk of developing HZ. \citet{yang2011risk}
find a positive association between COPD and development of HZ; however,
they do not control for important counfounders between COPD and HZ,
for example, cigarette smoking and alcohol consumption.

We analyze the main data from the 2005 Longitudinal Health Insurance
Database (LHID, \citealp{yang2011risk}) and the validation data from
the 2005 National Health Interview Survey conducted by the National
Health Research Institute and the Bureau of Health Promotion in Taiwan
\citep{lin2014adjustment}. The 2005 LHID consist of $42,430$ subjects
followed from the date of cohort entry on January 1, 2004 until the
development of HZ or December 31, 2006, whichever came first. Among
those, $8,486$ subjects were having COPD, denoted by $A=1$, and
$33,944$ subjects were not, denoted by $A=0$. The outcome $Y$ was
the development of HZ during follow up ($1$, having HZ and $0$,
not having HZ). The observed prevalence of HZ among COPD and non-COPD
subjects are $3.7\%$ and $2.2\%$ in the main data and $2.5\%$ and
$0.8\%$ in the validation data.

The confounders $X$ available from the main data were age, sex, diabetes
mellitus, hypertension, coronary artery disease, chronic liver disease,
autoimmune disease, and cancer. However, important confounders $U$,
including cigarette smoking and alcohol consumption, were not available.
The validation data $\mathcal{O}_{2}$ use the same inclusion criteria
as in the main study and consist of $1,148$ subjects who were comparable
to the subjects in the main data. Among those, $244$ subjects were
diagnosed of COPD, and $904$ subjects were not. In addition to all
variables available from the main data, cigarette smoking and alcohol
consumption were measured. In our formulation, the main data $\mathcal{O}_{1}$
combine the LHID data and the validation data. Table 4 in \citet{lin2014adjustment}
shows summary statistics on demographic characteristics and comorbid
disorders for COPD and Non-COPD subjects in the main and validation
data. Because the common covariates in the main and validation data
are comparable, it is reasonable to assume that the validation sample
is a simple random sample from the main data. Moreover, the difference
in distributions of alcohol consumption between COPD and non-COPD
subjects is not statistical significant in the validation data. But,
the COPD subjects tended to have higher cumulative smoking rates than
the non-COPD subjects in the validation data.

We obtain the initial estimators applied solely to the validation
data and the proposed estimators applied to both data. As suggested
by the simulation in Section \ref{sec:Simulation}, we use the same
type of the error-prone estimator as the initial estimator. Following
\citet{sturmer2005adjusting} and \citet{lin2014adjustment}, we use
the propensity score to accommodate the high-dimensional confounders.
Specifically, we fit logistic regression models for the propensity
score $e(X,U;\alpha)$ and the error-prone propensity score $e(X;\gamma)$
based on $\{(A_{j},X_{j},U_{j}):j\in\text{\ensuremath{\mathcal{S}}}_{2}\}$
and $\{(A_{i},X_{i}):j\in\mathcal{S}_{1}\}$, respectively. We fit
logistic regression models for the outcome mean function $\mu_{a}(X,U)$
based on a linear predictor $\{1,e(X,U;\hat{\alpha})\}^{\T}\beta_{a}$,
and for $\mu_{a}(X)$ based on a linear predictor $\{1,e(X;\hat{\gamma})\}^{\T}\eta_{a}$,
for $a=0,1$.

We first estimate the ACE $\tau.$ Table \ref{tab:Results-2} shows
the results for the average COPD effect on the development of HZ.
We find no big differences in the point estimates between our proposed
estimators and the corresponding initial estimators, but large reductions
in the estimated standard errors of the proposed estimators. As a
result, all $95\%$ confidence intervals based on the initial estimators
include $0$, but the $95\%$ confidence intervals based on the proposed
estimators do not include $0$, except for $\hat{\tau}_{\mat2\&\mat}$.
As demonstrated by the simulation in Section \ref{sec:Simulation},
the variance reduction by utilizing the main data is the smallest
for the matching estimator. From the results, on average, COPD increases
the percentage of developing HZ by $1.55\%$.

We also estimate the log of the causal risk ratio of HZ with COPD.
The initial IPW estimate from the validation data is $\log\widehat{\text{CRR}}_{\ipw,2}$$=1.10$
($95\%$ confidence interval: $0.02,\ 2.18$). In contrast, the proposed
estimate by using the error-prone IPW estimators is $\log\widehat{\text{CRR}}_{\ipw,2\&\ipw}$$=0.57$
($95\%$ confidence interval: $0.41,\ 0.72$), which is much more
accurate than the initial IPW estimate.

\begin{table}[H]
\caption{\label{tab:Results-2}Point estimate, bootstrap standard error 
 and $95\%$ confidence interval}

\centering{}{\small{}{}{}{}{}{}{}{}{}{}}%
\begin{tabular}{lccc|lccc}
\hline 
 & {\small{}{}Est}  & {\small{}{}SE}  & {\small{}{}{}{}{}{}{}{}{}{}$95\%$ CI}  &  & {\small{}{}Est}  & {\small{}{}{}{}{}{}{}{}{}{}SE}  & {\small{}{}{}{}{}{}{}{}{}{}$95\%$ CI}\tabularnewline
\hline 
{\small{}{}$\hat{\tau}_{\reg,2}$}  & {\small{}{}$0.0178$}  & {\small{}{}$0.0112$}  & {\small{}{}$(-0.0047,0.0402)$}  & {\small{}{}$\hat{\tau}_{\reg,2\&\reg}$ }  & {\small{}{}$0.0155$}  & {\small{}{}{}{}{}{}{}{}{}{}$0.0023$}  & {\small{}{}{}{}{}{}{}{}{}{}$(0.0109,0.0200)$} \tabularnewline
{\small{}{}$\hat{\tau}_{\ipw,2}$}  & {\small{}{}$0.0175$}  & {\small{}{}$0.0111$}  & {\small{}{}$(-0.0048,0.0398)$}  & {\small{}{}$\hat{\tau}_{\ipw,2\&\ipw}$ }  & {\small{}{}$0.0155$}  & {\small{}{}{}{}{}{}{}{}{}{}$0.0024$}  & {\small{}{}{}{}{}{}{}{}{}{}$(0.0108,0.0202)$} \tabularnewline
{\small{}{}$\hat{\tau}_{\aipw,2}$}  & {\small{}{}$0.0179$}  & {\small{}{}$0.0111$}  & {\small{}{}$(-0.0044,0.0402)$}  & {\small{}{}$\hat{\tau}_{\aipw,2\&\aipw}$}  & {\small{}{}$0.0156$}  & {\small{}{}{}{}{}{}{}{}{}{}$0.0024$}  & {\small{}{}{}{}{}{}{}{}{}{}$(0.0109,0.0203)$}\tabularnewline
{\small{}{}$\hat{\tau}_{\mat,2}$}  & {\small{}{}$0.0077$}  & {\small{}{}$0.0092$}  & {\small{}{}$(-0.0106,0.021)$}  & {\small{}{}$\hat{\tau}_{\mat,2\&\mat}$}  & {\small{}{}$0.0079$}  & {\small{}{}{}{}{}{}{}{}{}{}$0.0053$}  & {\small{}{}{}{}{}{}{}{}{}{}$(-0.0027,0.0184)$}\tabularnewline
\hline 
\end{tabular}
\end{table}

\section{Relaxing Assumption \ref{assump: m}\label{sec:Another-perspective}}

\textcolor{black}{{} In previous sections, we invoked Assumption \ref{assump: m}
that $\mathcal{S}_{2}$ is a random sample from $\mathcal{S}_{1}$.
We now relax this assumption and link our framework to existing methods
for missing data. Let $I_{i}$ be the indicator of selecting unit
$i$ into the validation data, i.e., $I_{i}=1$ if $i\in\mathcal{S}_{2}$
and $I_{i}=0$ if $i\notin\mathcal{S}_{2}$. Alternatively, $I_{i}$
can be viewed as the missingness indicator of $U_{i}$. Under Assumption
\ref{assump: m}, $I\indep(A,X,U,Y)$; i.e., $U$ is missing completely
at random. We now relax it to $I\indep U\mid(A,X,Y)$, i.e., $U$
is missing at random. In this case, the selection of $\mathcal{S}_{2}$
from $\mathcal{S}_{1}$ can depend on a probability design, which
is common in observational studies, e.g., an outcome-dependent two-phase
sampling \citep{breslow2003large,wang2009causal}.}

We assume that each unit in the main data is subjected to an independent
Bernoulli trial which determines whether the unit is selected into
the validation data. For simplicity, we further assume that the inclusion
probability $\pr(I=1\mid A,X,U,Y)=\pr(I=1\mid A,X,Y)\equiv\pi(A,X,Y)$
is known as in two-phase sampling. Otherwise, we need to fit a model
for the missing data indicator $I$ given $(A,X,Y)$. We summarize
the above in the following assumption.

\begin{assumption}\label{asump-missing} $\{(I_{i},A_{i},X_{i},U_{i},Y_{i}):i\in\mathcal{S}_{1}\}$
are IID with $I\indep U\mid(A,X,Y)$. $\mathcal{S}_{2}$ is selected
from $\mathcal{S}_{1}$ with a known inclusion probability $\pi(A,X,Y)>0$.
\end{assumption}

In what follows, we use $\pi$ for $\pi(A,X,Y)$ and $\pi_{j}$ for
$\pi(A_{j},X_{j},Y_{j})$ for shorthand. Because of Assumption \ref{asump-missing},
we drop the indices $i$ and $j$ in the expectations, covariances,
and variances, which are taken with respect to both the sampling and
superpopulation models.

\subsection{RAL estimators\label{subsec:RelaxationRAL}}

For the illustration of RAL estimators, we focus on the AIPW estimator
of the ACE $\tau$, because the regression imputation and inverse
probability weighting estimators are its special cases. Let $\hat{\alpha}$
and $\hat{\beta}_{a}$ solve the weighted estimating equations $\sum_{j\in\mathcal{S}_{2}}\pi_{j}^{-1}S(A_{j},X_{j},U_{j};\alpha)=0$
and $\sum_{j\in\mathcal{S}_{2}}\pi_{j}^{-1}S_{a}(A_{j},X_{j},U_{j},Y_{j};\beta_{a})=0$,
and let $\alpha^{*}$ and $\beta_{a}^{*}$ satisfy $E\{S(A,X,U;\alpha^{*})\}=0$
and $E\{S_{a}(A,X,U,Y;\beta_{a}^{*})\}=0$. Under suitable regularity
condition, $\hat{\alpha}\rightarrow\alpha^{*}$ and $\hat{\beta}_{a}\rightarrow\beta_{a}^{*}$
in probability, for $a=0,1$. Let the initial estimator for $\tau$
be the Hajek-type estimator \citep{hajek1971comment}: 
\begin{equation}
\hat{\tau}_{2}=\frac{\sum_{j\in\mathcal{S}_{2}}\pi_{j}^{-1}\hat{\tau}_{\aipw,2,j}}{\sum_{j\in\mathcal{S}_{2}}\pi_{j}^{-1}},\label{eq:Hajek}
\end{equation}
where $\hat{\tau}_{\aipw,2,j}$ has the same form as (\ref{eq::aipw-ind}).
Under regularity conditions, Assumption \ref{asump outcome} or \ref{asump ps},
and Assumption \ref{asump-missing}, we show in the supplementary
material that\textcolor{black}{{} 
\begin{equation}
\hat{\tau}_{2}-\tau\cong n_{1}^{-1}\sum_{j\in\mathcal{S}_{1}}\pi_{j}^{-1}I_{j}\psi(A_{j},X_{j},U_{j},Y_{j}),\label{eq:IF_for_RAL2}
\end{equation}
}where $\psi(A,X,U,Y)$ is given by (\ref{eq:aipw influence fctn}).
Because \textcolor{black}{$\{\pi_{j}^{-1}I_{j}\psi(A_{j},X_{j},U_{j},Y_{j}):j\in\mathcal{S}_{1}\}$
are IID with mean $0$, $\hat{\tau}_{2}$ is consistent for $\tau$.}

Similarly, let $\hat{\gamma}_{d}$ and $\hat{\eta}_{d,a}$ solve the
weighted estimating equation $\sum_{j\in\mathcal{S}_{d}}\pi_{j}^{-1}S(A_{j},X_{j};\gamma)=0$
and $\sum_{j\in\mathcal{S}_{d}}\pi_{j}^{-1}S_{a}(A_{j},X_{j},Y_{j};\eta_{a})=0,$
and let $\gamma^{*}$ and $\eta_{a}^{*}$ satisfy $E\{S(A_{j},X_{j};\gamma^{*})\}=0$
and $E\{S_{a}(A_{j},X_{j},Y_{j};\eta_{a}^{*})\}=0$ . Under suitable
regularity condition, $\hat{\gamma}_{d}\rightarrow\gamma^{*}$ and
$\hat{\eta}_{d,a}\rightarrow\eta_{a}^{*}$ in probability, for $a=0,1$
and $d=1,2$. Let the error-prone estimators be 
\begin{equation}
\hat{\tau}_{1,\ep}=n_{1}^{-1}\sum_{i\in\mathcal{S}_{1}}\hat{\tau}_{\aipw,1,\ep,i},\qquad\hat{\tau}_{2,\ep}=\frac{\sum_{j\in\mathcal{S}_{2}}\pi_{j}^{-1}\hat{\tau}_{\aipw,2,\ep,j}}{\sum_{j\in\mathcal{S}_{2}}\pi_{j}^{-1}},\label{eq:Hajek2}
\end{equation}
where $\hat{\tau}_{\aipw,d,\ep,j}$ has the same form as (\ref{eq:aipw ep j})
in the supplementary material. Following a similar derivation for
(\ref{eq:IF_for_RAL2}), we have\textcolor{black}{{} 
\begin{equation}
\hat{\tau}_{1,\ep}-\tau_{\ep}\cong n_{1}^{-1}\sum_{i\in\mathcal{S}_{1}}\phi(A_{i},X_{i},Y_{i}),\quad\hat{\tau}_{2,\ep}-\tau_{\ep}\cong n_{1}^{-1}\sum_{j\in\mathcal{S}_{1}}\pi_{j}^{-1}I_{j}\phi(A_{j},X_{j},Y_{j}),\label{eq:IF_for_RALep}
\end{equation}
}where $\phi(A,X,Y)$ is given by (\ref{eq:ep-aipw}) in the supplementary
material. Because \textcolor{black}{both $\{\phi(A_{i},X_{i},Y_{i}):i\in\mathcal{S}_{1}\}$
and $\{\pi_{j}^{-1}I_{j}\phi(A_{j},X_{j},Y_{j}):j\in\mathcal{S}_{1}\}$
are IID with mean $0$, $\hat{\tau}_{1,\ep}$ and $\hat{\tau}_{2,\ep}$
are consistent for $\tau_{\ep}$.}

\begin{theorem}\label{Thm:relaxRAL}Under certain regularity conditions,
(\ref{eq:asymp}) holds for the Hajek-type estimators (\ref{eq:Hajek})
and (\ref{eq:Hajek2}), where $\rho=\plim_{n_{2}\rightarrow\infty}\left(n_{2}/n_{1}\right)$,
$v_{2}=\rho\times\var\{\pi^{-1}I\psi(A,X,U,Y)\}$, $\Gamma=\rho\times\cov\{\pi^{-1}I\psi(A,X,$
$U,Y),(\pi^{-1}I-1)\phi(A,X,Y)\}$ and $V=\rho\times\var\{(\pi^{-1}I-1)\phi(A,X,Y)\}$.

\end{theorem}

Similar to Section \ref{subsec:RAL}, we can construct a consistent
variance estimator for $\hat{\tau}$ by replacing the variances and
covariance in Theorem \ref{Thm:relaxRAL} with their sample analogs.

\subsection{Matching estimators\label{subsec:Relaxation-mat}}

Recall that $\J_{d,V,l}$ is the index set of matches for unit $l$
based on data $\mathcal{O}_{d}$ and the matching variable $V$, which
can be $(X,U)$ or $X$. Define $\delta_{d,V,(j,l)}=1$ if $j\in\J_{d,V,l}$
and $\delta_{d,V,(j,l)}=0$ otherwise. Now, we denote $K_{d,V,j}=\pi_{j}\sum_{l\in\mathcal{S}_{d}}\pi_{l}^{-1}1\{A_{l}=1-A_{j}\}\delta_{d,V,(j,l)}$
as the weighted number of times that unit $j$ is used as a match.
If $\pi_{j}$ is a constant for all $j\in\mathcal{S}_{d}$, then $K_{d,V,j}$
reduces to the number of times that unit $j$ is used as a match defined
in Section \ref{sec::estimators}, which justifies using the same
notation as before.

Let the initial matching estimator for $\tau$ be the Hajek-type estimator:

\begin{eqnarray*}
\hat{\tau}_{\mat,2}^{(0)} & = & \left(\sum_{j\in\mathcal{S}_{2}}\pi_{j}^{-1}\right)^{-1}\sum_{j\in\mathcal{S}_{2}}\pi_{j}^{-1}(2A_{j}-1)\left(Y_{j}-M^{-1}\sum_{l\in\J_{2,(X,U),j}}Y_{l}\right)\\
 & = & \left(\sum_{j\in\mathcal{S}_{2}}\pi_{j}^{-1}\right)^{-1}\sum_{j\in\mathcal{S}_{2}}\pi_{j}^{-1}(2A_{j}-1)\left\{ 1+M^{-1}K_{2,(X,U),j}\right\} Y_{j}.
\end{eqnarray*}
Let a bias-corrected matching estimator be 
\begin{equation}
\hat{\tau}_{\mat,2}=\hat{\tau}_{\mat,2}^{(0)}-n_{1}^{-1/2}\hat{B}_{2},\label{eq:mat2 new}
\end{equation}
where 
\[
\hat{B}_{2}=n_{1}^{-1/2}\sum_{j\in\mathcal{S}_{2}}\pi_{j}^{-1}(2A_{j}-1)\left[M^{-1}\sum_{l\in\J_{2,(X,U),j}}\left\{ \hat{\mu}_{1-A_{j}}(X_{j},U_{j})-\hat{\mu}_{1-A_{j}}(X_{l},U_{l})\right\} \right],
\]
We show in the supplementary material that 
\begin{equation}
\hat{\tau}_{\mat,2}-\tau\cong n_{1}^{-1}\sum_{j\in\mathcal{S}_{2}}\pi_{j}^{-1}\psi_{\mat,j},\label{eq:pi-matching}
\end{equation}
where $\psi_{\mat,j}$ is defined in (\ref{eq:psi_mat}) with the
new definition of $K_{2,(X,U),j}$.

Similarly, we obtain error-prone matching estimators and express them
as 
\begin{equation}
\hat{\tau}_{\mat,1,\ep}-\tau_{\ep}\cong n_{1}^{-1}\sum_{j\in\mathcal{S}_{1}}\phi_{\mat,1,j},\quad\hat{\tau}_{\mat,2,\ep}-\tau_{\ep}\cong n_{1}^{-1}\sum_{j\in\mathcal{S}_{2}}\pi_{j}^{-1}\phi_{\mat,2,j},\label{eq:pi-matching ep}
\end{equation}
where $\phi_{\mat,d,j}$ is defined in (\ref{eq:phi_mat}) with the
new definition of $K_{d,X,j}$.

From the above decompositions, $\hat{\tau}_{\mat,2}$ is consistent
for $\tau$, and $\hat{\tau}_{\mat,1,\ep}-\hat{\tau}_{\mat,2,\ep}$
is consistent for $0$.

\begin{theorem}\label{Thm:relaxMat}Under certain regularity conditions,
(\ref{eq:asymp}) holds for the estimators (\ref{eq:mat2 new}) and
$\hat{\tau}_{\mat,d,\ep}$ ($d=1,2$), where \textcolor{black}{$\rho=\plim_{n_{2}\rightarrow\infty}\left(n_{2}/n_{1}\right)$,}
\begin{eqnarray*}
v_{2} & = & \rho\times\left(E\left[\frac{1-\pi}{\pi}\left\{ \tau(X,U)-\tau\right\} ^{2}\right]\right.\\
 &  & +\left.\plim\left[n_{1}^{-1}\sum_{j\in\mathcal{S}_{1}}\frac{1-\pi_{j}}{\pi_{j}}\left\{ 1+M^{-1}K_{2,(X,U),j}\right\} ^{2}\sigma_{A_{j}}^{2}(X_{j},U_{j})\right]\right),\\
\Gamma & = & \rho\times E\left[\frac{1-\pi}{\pi}\left\{ \mu_{1}(X,U)-\mu_{0}(X,U)-\tau\right\} \left\{ \mu_{1}(X)-\mu_{0}(X)-\tau_{\ep}\right\} \right]\\
 &  & +\rho\times\plim\left[n_{1}^{-1}\sum_{j\in\mathcal{S}_{1}}\frac{1-\pi_{j}}{\pi_{j}}\left\{ 1+M^{-1}K_{2,(X,U),j}\right\} \left(1+M^{-1}K_{2,X,j}\right)\sigma_{A_{j}}^{2}(X_{j},U_{j})\right],\\
V & = & \rho\times E\left[\frac{1-\pi}{\pi}\left\{ \mu_{1}(X)-\mu_{0}(X)-\tau_{\ep}\right\} ^{2}\right]\\
 &  & +\rho\times\plim\left[n_{1}^{-1}\sum_{j\in S_{1}}\frac{1-\pi_{j}}{\pi_{j}}\left(1+M^{-1}K_{1,X,j}\right)^{2}\sigma_{A_{j}}^{2}(X_{j})\right].
\end{eqnarray*}

\end{theorem}

We can construct variance estimators based on the formulas in Theorem
\ref{Thm:relaxMat}. However, this will again involve estimating the
conditional variances $\sigma_{0}^{2}(x)$ and $\sigma_{1}^{2}(x)$.
We recommend using a bootstrap variance estimator proposed in the
next subsection.

\subsection{A bootstrap variance estimation procedure}

The asymptotic linear forms (\ref{eq:IF_for_RAL2}), (\ref{eq:IF_for_RALep}),
(\ref{eq:pi-matching}), and (\ref{eq:pi-matching ep}) are useful
for the bootstrap variance estimation. For $b=1,\ldots,B$, we construct
the bootstrap replicates fas follows: 
\begin{description}
\item [{Step$\ 1.$}] Sample $n_{1}$ units from $\mathcal{S}_{1}$ with
replacement as $\mathcal{S}_{1}^{*(b)}$. 
\item [{Step$\ 2.$}] Compute the bootstrap replicates of $\hat{\tau}_{2}-\tau$
and $\hat{\tau}_{d,\ep}-\tau_{\ep}$ as 
\begin{eqnarray*}
\hat{\tau}_{2}^{(b)}-\hat{\tau}_{2} & = & n_{1}^{-1}\sum_{i\in\mathcal{S}_{1}^{*(b)}}\pi_{i}^{-1}I_{i}\hat{\psi}_{i},\\
\hat{\tau}_{1,\ep}^{(b)}-\hat{\tau}_{1,\ep} & = & n_{1}^{-1}\sum_{i\in\mathcal{S}_{1}^{*(b)}}\hat{\phi}_{1,i},\\
\hat{\tau}_{2,\ep}^{(b)}-\hat{\tau}_{2,\ep} & = & n_{1}^{-1}\sum_{i\in\mathcal{S}_{1}^{*(b)}}\pi_{i}^{-1}I_{i}\hat{\phi}_{2,i},
\end{eqnarray*}
where $(\hat{\psi}_{i},\hat{\phi}_{d,i})$ are the estimated versions
of $(\psi_{i},\phi_{i})$ from $\mathcal{O}_{d}$ ($d=1,2$). 
\end{description}
We estimate $\Gamma,V$ and $v_{2}$ by (\ref{eq:gamma-hat})--(\ref{eq:v2-hat})
based on the above bootstrap replicates, and $\var(\hat{\tau})$ by
(\ref{eq:VE}), i.e., $\hat{v}=\hat{v}_{2}-\hat{\Gamma}^{\T}\hat{V}^{-1}\hat{\Gamma}.$

\begin{theorem}\label{Thm: boot-1}Under certain regularity conditions,
$(\hat{\Gamma},\hat{V},\hat{v}_{2},\hat{v})$ are consistent for $\{\Gamma,V,\var(\hat{\tau}_{2}),\var(\hat{\tau})\}$.

\end{theorem}

For RAL estimators, we can also use the classical nonparametric bootstrap
based on resampling the IID observations $\{(I_{i},A_{i},X_{i},U_{i},Y_{i}):i\in\mathcal{S}_{1}\}$
and repeating the analysis as for the original data. The above bootstrap
procedure based on resampling the linear forms are particularly useful
for the matching estimator.

\subsection{Connection with missing data}

As a final remark, we express the proposed estimator in a linear form
that has appeared in the missing data literature.

\begin{proposition}\label{prop1}Under certain regularity conditions
and Assumption \ref{asump-missing}, $\hat{\tau}$ has an asymptotic
linear form 
\begin{equation}
n_{1}^{1/2}(\hat{\tau}-\tau)=n_{1}^{-1/2}\sum_{i\in\mathcal{S}_{1}}\left\{ \frac{I_{i}}{\pi_{i}}\psi_{i}-\left(\frac{I_{i}}{\pi_{i}}-1\right)\Gamma V^{-1}\phi_{i}\right\} +o_{P}(1),\label{eq:asymp linear}
\end{equation}
where $\psi_{i}$ is $\psi(A_{i},X_{i},U_{i},Y_{i})$ for RAL estimators
and $\psi_{\mat,i}$ for the matching estimator, and a similar definition
applies to $\phi_{i}$. Under Assumption \ref{assump: m}, $\pi_{i}\equiv\rho$.

\end{proposition}

Expression (\ref{eq:asymp linear}) is within a class of estimators
in the missing data literature with the form 
\begin{equation}
n_{1}^{1/2}(\hat{\tau}-\tau)=n_{1}^{-1/2}\sum_{i\in\mathcal{S}_{1}}\left\{ \frac{I_{i}}{\pi_{i}}s(A_{i},X_{i},U_{i},Y_{i})-\left(\frac{I_{i}}{\pi_{i}}-1\right)\text{\ensuremath{\kappa}}(A_{i},X_{i},Y_{i})\right\} +o_{P}(1),\label{eq:missing score}
\end{equation}
where $\pi=E(I\mid A,X,U,Y)$, $s(A,X,U,Y)$ satisfies $E\{s(A,X,U,Y)\}=0$,
and $s(A,X,U,Y)$ and $\text{\ensuremath{\kappa}}(A,X,Y)$ are square
integrable. Given $s(A,X,U,Y)$, the optimal choice of $\text{\ensuremath{\kappa}}(A,X,Y)$
is $\kappa_{\opt}(A,X,Y)=E\{s(A,X,U,Y)\mid A,X,Y\}$, which minimizes
the asymptotic variance of (\ref{eq:missing score}) \citep{robins1994estimation,wang2009causal}.
However, $\kappa_{\opt}(A,X,Y)$ requires a correct specification
of the missing data model $f(U\mid A,X,Y)$. In our approach, instead
of specifying the missing data model, we specify the error-prone estimators
and utilize an estimator that is consistent for zero to improve the
efficiency of the initial estimator. This is more attractive and closer
to empirical practice than calculating $\kappa_{\opt}(A,X,Y)$, because
practitioners only need to apply their favorite estimators to the
main and validation data using existing software. See also \citet{chen2000unified}
for a similar discussion in the regression context under Assumption
\ref{assump: m}.

\section{Discussion}

Depending on the roles in statistical inference, there are two types
of \textit{big data}: one with large-sample sizes and the other with
richer covariates. In our discussion, the main data have a larger
sample size, and the validation data have more covariates. Although
some counterexamples exist \citep{pearl2009letter,pearl2010class},
it is more reliable to draw causal inference from the validation data.
The proposed strategy is applicable even the number of covariates
is high or the sample size is small in the validation data. In this
case, we can consider $\hat{\tau}_{2}$ to be the double machine learning
estimators \citep{chernozhukov2018double} that use flexible machine
learning methods for estimating regression and propensity score functions
while retain the property in (\ref{eq:asymp}). Our framework allows
for more efficient estimators of the causal effects by further combining
information in the main data, without imposing any parametric models
for the partially observed covariates. Coupled with the bootstrap,
our estimators require only software implementations of standard estimators,
and thus are attractive for practitioners who want to combine multiple
observational data sources.

The key insight is to leverage an estimator of zero to improve the
efficiency of the initial estimator. If a certain feature is transportable,
we can easily construct an estimator of zero. For example. We have
shown that if the validation sample is a simple random sample from
the main sample, the distribution of $(A,X,Y)$ is transportable from
the validation sample to the main sample. We can construct an estimator
of zero by taking the difference of the estimators based on $(A,X,Y)$
from the two data. In the presence of heterogeneity between two samples,
the transportability of the whole distribution of $(A,X,Y)$ can be
too strong. However, if we are willing to assume the conditional distribution
of $Y$ given $(A,X)$ is transportable, we can then take the error
prone estimators to be the regression coefficients of $Y$ on $(A,X)$
from the two datasets. As suggested by one of the reviewers, if the
subgroups of two samples are comparable, we can construct the error
prone estimators based on the subgroups. Similarly, this construction
of error prone estimators can adopt to different transportability
assumptions based on the subject matter knowledge.

In the worse case, the heterogeneity is intrinsic between the two
samples, and we cannot construct two error prone estimators with the
same limit. We can still conduct a sensitivity analysis combining
two data. Instead of (\ref{eq:asymp}), we assume 
\begin{equation}
n_{2}^{1/2}\left(\begin{array}{c}
\hat{\tau}_{2}-\tau\\
\hat{\tau}_{2,\ep}-\hat{\tau}_{1,\ep}-\delta
\end{array}\right)\rightarrow\N\left\{ 0_{L+1},\left(\begin{array}{cc}
v_{2} & \Gamma^{\T}\\
\Gamma & V
\end{array}\right)\right\} ,\label{eq:new asymp}
\end{equation}
where $\delta$ is the sensitivity parameter, quantifying the systematic
difference between $\hat{\tau}_{2,\ep}$ and $\hat{\tau}_{1,\ep}$.
The adjusted estimator becomes $\hat{\tau}_{\mathrm{adj}}(\delta)=\hat{\tau}_{2}-\hat{\Gamma}^{\T}\hat{V}^{-1}(\hat{\tau}_{2,\ep}-\hat{\tau}_{1,\ep}-\delta).$
With different values of $\delta$, the estimator $\hat{\tau}_{\mathrm{adj}}(\delta)$
can provide valuable insight on the impact of the heterogeneity of
the two data, allowing an investigator to assess the extent to which
the heterogeneity may alter causal inferences about treatment effects.

\section*{Acknowledgments}

We thank the editor, the associate editor, and four anonymous reviewers
for insightful comments and suggestions which improved the article
significantly. We are grateful to Professor Yi-Hau Chen for providing
the data used in this paper and for offering help and advice in interpreting
the data and Drs. Lo-Hua Yuan and Xinran Li for helpful comments.
Dr. Yang is partially supported by the National Science Foundation
grant DMS 1811245, National Cancer Institute grant P01 CA142538, and
Oak Ridge Associated Universities. Dr. Ding is partially supported
by the National Science Foundation grant DMS 1713152.

\section*{Supplementary materials}

The online supplementary material contains technical details and proofs,
and the R package "IntegrativeCI" is available at \url{https://github.com/shuyang1987/IntegrativeCI}
to perform the proposed estimators.

\bibliographystyle{dcu}
\bibliography{cipeng}

\pagebreak{}
\begin{center}
\textbf{\LARGE{}Supplementary materials for ``Combining multiple observational
data sources to estimate causal effects''}{\LARGE\par}
\par\end{center}

\medskip{}

\bigskip{}
\global\long\def\theequation{S\arabic{equation}}%
 \setcounter{equation}{0}

\global\long\def\thelemma{S\arabic{lemma}}%
 \setcounter{lemma}{0}

\global\long\def\theexample{S\arabic{example}}%
 \setcounter{equation}{0}

\global\long\def\thesection{S\arabic{section}}%
 \setcounter{section}{0}

\global\long\def\thetheorem{S\arabic{theorem}}%
 \setcounter{equation}{0}

\global\long\def\thecondition{S\arabic{condition}}%
 \setcounter{equation}{0}

\global\long\def\theremark{S\arabic{remark}}%
 \setcounter{equation}{0}

\global\long\def\thestep{S\arabic{step}}%
 \setcounter{equation}{0}

\global\long\def\theassumption{S\arabic{assumption}}%
 \setcounter{assumption}{0}

\global\long\def\theproof{S\arabic{proof}}%
 \setcounter{equation}{0}

\global\long\def\theproposition{S{proposition}}%
 \setcounter{equation}{0}

\section{Proof for Theorem \ref{Thm for RAL}}

The asymptotic normality holds for $n_{2}^{1/2}(\hat{\tau}_{2}-\tau,\hat{\tau}_{2,\ep}-\hat{\tau}_{1,\ep})$
by the moment conditions for the RAL estimators and the central limit
theorem. We then show the asymptotic variance formula in (\ref{eq:asymp}).

Based on the linear form in (\ref{eq:RAL1}), $v_{2}=\var\left\{ \psi(A,X,U,Y)\right\} $.
Based on the linear form in (\ref{eq:RAL2}), 
\begin{equation}
n_{2}^{1/2}(\hat{\tau}_{2,\ep}-\hat{\tau}_{1,\ep})\cong n_{2}^{-1/2}\sum_{j\in\mathcal{S}_{2}}\phi(A_{j},X_{j},Y_{j})-\rho^{1/2}n_{1}^{-1/2}\sum_{i\in\mathcal{S}_{1}}\phi(A_{i},X_{i},Y_{i}).\label{eq:Vep1}
\end{equation}
The first term in (\ref{eq:Vep1}) contribute $\var\{\phi(A,X,Y)\}$,
the second term contribute $\rho\var\{\phi(A,X,Y)\}$, and their correlation
contribute $-2\rho^{1/2}\rho^{1/2}\var\{\phi(A,X,Y)\}$. Therefore,
\begin{eqnarray*}
V & = & \var\{\phi(A,X,Y)\}+\rho\var\{\phi(A,X,Y)\}-2\rho^{1/2}\rho^{1/2}\var\{\phi(A,X,Y)\}\\
 & = & (1-\rho)\times\var\{\phi(A,X,Y)\}.
\end{eqnarray*}

To obtain the expression for $\Gamma,$ we re-write 
\[
n_{2}^{1/2}(\hat{\tau}_{2,\ep}-\hat{\tau}_{1,\ep})\cong n_{2}^{-1/2}\sum_{j\in\mathcal{S}_{2}}(1-\rho)\phi(A_{j},X_{j},Y_{j})-\rho n_{2}^{-1/2}\sum_{i\in\mathcal{S}_{1}\backslash\mathcal{S}_{2}}\phi(A_{i},X_{i},Y_{i}).
\]
Because observations in $\mathcal{S}_{2}$ and $\mathcal{S}_{1}\backslash\mathcal{S}_{2}$
are independent, simple calculations give 
\[
\Gamma=(1-\rho)\cov\{\psi(A,X,U,Y),\phi(A,X,Y)\}.
\]

\section{Proof of Remark \ref{rmk:best}}

For a smooth function $f(x,y,z)$, let $\partial f(x,y,z)/\partial x$,
$\partial f(x,y,z)/\partial y$ and $\partial f(x,y,z)/\partial z$
be its partial derivatives. By the Taylor expansion, we have 
\begin{eqnarray}
\hat{\tau} & = & f(\hat{\tau}_{2},\hat{\tau}_{1,\ep},\hat{\tau}_{2,\ep})\nonumber \\
 & \cong & f(\tau,\tau_{\ep},\tau_{\ep})+\frac{\partial f}{\partial x}(\tau,\tau_{\ep},\tau_{\ep})(\hat{\tau}_{2}-\tau)\nonumber \\
 &  & +\frac{\partial f}{\partial y^{\T}}(\tau,\tau_{\ep},\tau_{\ep})(\hat{\tau}_{1,\ep}-\tau_{\ep})+\frac{\partial f}{\partial z^{\T}}(\tau,\tau_{\ep},\tau_{\ep})(\hat{\tau}_{2,\ep}-\tau_{\ep})\nonumber \\
 & \equiv & l_{0}+l_{1}\hat{\tau}_{2}+l_{2}^{\T}\hat{\tau}_{1,\ep}+l_{3}^{\T}\hat{\tau}_{2,\ep},\label{eq:line2}
\end{eqnarray}
where 
\begin{eqnarray*}
l_{0} & = & f(\tau,\tau_{\ep},\tau_{\ep})-\frac{\partial f(\tau,\tau_{\ep},\tau_{\ep})}{\partial x}\tau-\left\{ \frac{\partial f(\tau,\tau_{\ep},\tau_{\ep})}{\partial y^{\T}}+\frac{\partial f(\tau,\tau_{\ep},\tau_{\ep})}{\partial z^{\T}}\right\} \tau_{\ep},\\
l_{1} & = & \frac{\partial f(\tau,\tau_{\ep},\tau_{\ep})}{\partial x},\quad l_{2}=\frac{\partial f(\tau,\tau_{\ep},\tau_{\ep})}{\partial y},\quad l_{3}=\frac{\partial f(\tau,\tau_{\ep},\tau_{\ep})}{\partial z}.
\end{eqnarray*}
Because $\hat{\tau}$ is consistent for $\tau$, letting $n_{2}$
go to infinity in (\ref{eq:line2}), we have $\tau=l_{0}+l_{1}\tau+l_{2}^{\T}\tau_{\ep}+l_{3}^{\T}\tau_{\ep}$
for all $\tau$ and $\tau_{\ep}.$ Then, it follows $l_{0}=0$, $l_{1}=1$,
and $l_{2}=-l_{3}$. Therefore, $\hat{\tau}\cong\hat{\tau}_{2}-l_{3}^{\T}(\hat{\tau}_{2,\ep}-\hat{\tau}_{1,\ep})$.
By the joint normality of $(\hat{\tau},\hat{\tau}_{1,\ep},\hat{\tau}_{2,\ep})$,
the optimal $l_{3}$ minimizing $\var\left\{ \hat{\tau}_{2}-l_{3}^{\T}(\hat{\tau}_{2,\ep}-\hat{\tau}_{1,\ep})\right\} $
is $\Gamma^{\T}V^{-1}$. Therefore, the proposed estimator in (\ref{eq:proposed estimator})
is optimal among the class of consistent estimators that is a smooth
function of $(\hat{\tau},\hat{\tau}_{1,\ep},\hat{\tau}_{2,\ep})$.

\section{Proof of Remark \ref{choice of ep}}

We can view $v_{2}-\Gamma^{\T}V^{-1}\Gamma$ and $\Gamma_{1}^{\T}V_{11}^{-1}\Gamma_{1}$
as the conditional variances in a multivariate normal vector. Then
the conclusion holds immediately because conditioning on more variables
will decrease the variance for a multivariate normal vector. For algebraic
completeness, we give a formal proof below.

Decompose the $L$-dimensional error-prone estimator $\hat{\tau}_{d,\ep}$
into two components with $L_{1}$ and $L_{2}$ dimensions respectively.
Then, $\Gamma$ and $V$ have the corresponding partitions: 
\[
\Gamma=\left(\begin{array}{c}
\Gamma_{1}\\
\Gamma_{2}
\end{array}\right),\quad V=\left(\begin{array}{cc}
V_{11} & V_{12}\\
V_{12}^{\T} & V_{22}
\end{array}\right).
\]
We assume $V$ is invertable, and therefore $V_{11}$ and $V_{22}$
are also invertable. To show that increasing the dimension of the
error-prone estimator would not decrease the efficiency gain, according
to (\ref{eq:asymp var}), it suffices to show that $\Gamma^{\T}V^{-1}\Gamma\geq\Gamma_{1}^{\T}V_{11}^{-1}\Gamma_{1}$.
Toward this end, note that 
\[
V^{-1}=\left(\begin{array}{cc}
V_{11}^{-1}+V_{11}^{-1}V_{12}(V_{22}-V_{12}^{\T}V_{11}^{-1}V_{12})^{-1}V_{12}^{\T}V_{11}^{-1} & -V_{11}^{-1}V_{12}(V_{22}-V_{12}^{\T}V_{11}^{-1}V_{12})^{-1}\\
-(V_{22}-V_{12}^{\T}V_{11}^{-1}V_{12})^{-1}V_{12}^{\T}V_{11}^{-1} & (V_{22}-V_{12}^{\T}V_{11}^{-1}V_{12})^{-1}
\end{array}\right),
\]
so that 
\begin{eqnarray}
\Gamma^{\T}V^{-1}\Gamma & = & \Gamma_{1}^{\T}V_{11}^{-1}\Gamma_{1}+\Gamma_{1}^{\T}\{V_{11}^{-1}V_{12}(V_{22}-V_{12}^{\T}V_{11}^{-1}V_{12})^{-1}V_{12}^{\T}V_{11}^{-1}\}\Gamma_{1}\nonumber \\
 &  & -\Gamma_{1}^{\T}V_{11}^{-1}V_{12}(V_{22}-V_{12}^{\T}V_{11}^{-1}V_{12})^{-1}\Gamma_{2}\nonumber \\
 &  & -\Gamma_{2}^{\T}(V_{22}-V_{12}^{\T}V_{11}^{-1}V_{12})^{-1}V_{12}^{\T}V_{11}^{-1}\Gamma_{1}\nonumber \\
 &  & +\Gamma_{2}^{\T}(V_{22}-V_{12}^{\T}V_{11}^{-1}V_{12})^{-1}\Gamma_{2}.\label{eq:8-1}
\end{eqnarray}
Let $\theta=(V_{22}-V_{12}^{\T}V_{11}^{-1}V_{12})^{-1/2}V_{12}^{\T}V_{11}^{-1}\Gamma_{1}$
and $\delta=(V_{22}-V_{12}^{\T}V_{11}^{-1}V_{12})^{-1/2}\Gamma_{2}$.
Then, (\ref{eq:8-1}) becomes 
\[
\Gamma^{\T}V^{-1}\Gamma=\Gamma_{1}^{\T}V_{11}^{-1}\Gamma_{1}+\theta^{\T}\theta-\theta^{\T}\delta-\delta^{\T}\theta+\delta^{\T}\delta=\Gamma_{1}^{\T}V_{11}^{-1}\Gamma_{1}+(\theta-\delta)^{\T}(\theta-\delta)\geq\Gamma_{1}^{\T}V_{11}^{-1}\Gamma_{1}.
\]

\section{Proof of Lemma \ref{lemma(AIPW)}\label{sec:Proof-of-Lemma 3 for aipw}}

We write $\hat{\tau}_{\aipw,2}=\hat{\tau}_{\aipw,2}(\hat{\alpha},\hat{\beta}_{0},\hat{\beta}_{1})$
to emphasize its dependence on the parameter estimates $(\hat{\alpha},\hat{\beta}_{0},\hat{\beta}_{1})$.
By the Taylor expansion, 
\begin{eqnarray}
 &  & \hat{\tau}_{\aipw,2}(\hat{\alpha},\hat{\beta}_{0},\hat{\beta}_{1})\nonumber \\
 & \cong & \hat{\tau}_{\aipw,2}(\alpha^{*},\beta_{0}^{*},\beta_{1}^{*})+E\left\{ \frac{\partial\hat{\tau}_{\aipw,2}(\alpha^{*},\beta_{0}^{*},\beta_{1}^{*})}{\partial\alpha^{\T}}\right\} (\hat{\alpha}-\alpha^{*})\nonumber \\
 &  & +E\left\{ \frac{\partial\hat{\tau}_{\aipw,2}(\alpha^{*},\beta_{0}^{*},\beta_{1}^{*})}{\partial\beta_{0}^{\T}}\right\} (\hat{\beta}_{0}-\beta_{0}^{*})+E\left\{ \frac{\partial\hat{\tau}_{\aipw,2}(\alpha^{*},\beta_{0}^{*},\beta_{1}^{*})}{\partial\beta_{1}^{\T}}\right\} (\hat{\beta}_{1}-\beta_{1}^{*})\nonumber \\
 & \cong & \hat{\tau}_{\aipw,2}(\alpha^{*},\beta_{0}^{*},\beta_{1}^{*})\nonumber \\
 &  & +n_{2}^{-1}\sum_{j\in\mathcal{S}_{2}}E\left\{ \frac{\partial\hat{\tau}_{\aipw,2}(\alpha^{*},\beta_{0}^{*},\beta_{1}^{*})}{\partial\alpha^{\T}}\right\} E\left\{ \frac{\partial S(A,X,U;\alpha^{*})}{\partial\alpha^{\T}}\right\} ^{-1}S(A_{j},X_{j},U_{j};\alpha^{*})\label{eq:line1}\\
 &  & -n_{2}^{-1}\sum_{j\in\mathcal{S}_{2}}E\left\{ \frac{\partial\hat{\tau}_{\aipw,2}(\alpha^{*},\beta_{0}^{*},\beta_{1}^{*})}{\partial\beta_{0}^{\T}}\right\} E\left\{ \frac{\partial S_{0}(A,X,U,Y;\beta_{0}^{*})}{\partial\beta_{0}^{\T}}\right\} ^{-1}S_{0}(A_{j},X_{j},U_{j},Y_{j};\beta_{0}^{*})\nonumber \\
 &  & -n_{2}^{-1}\sum_{j\in\mathcal{S}_{2}}E\left\{ \frac{\partial\hat{\tau}_{\aipw,2}(\alpha^{*},\beta_{0}^{*},\beta_{1}^{*})}{\partial\beta_{1}^{\T}}\right\} E\left\{ \frac{\partial S_{0}(A,X,U,Y;\beta_{0}^{*})}{\partial\beta_{1}^{\T}}\right\} ^{-1}S_{1}(A_{j},X_{j},U_{j},Y_{j};\beta_{1}^{*}).\nonumber 
\end{eqnarray}
We have the following calculations: 
\begin{eqnarray*}
E\left\{ \frac{\partial\hat{\tau}_{\aipw,2}(\alpha^{*},\beta_{0}^{*},\beta_{1}^{*})}{\partial\alpha}\right\}  & = & -H_{\aipw},\\
E\left\{ \frac{\partial\hat{\tau}_{\aipw,2}(\alpha^{*},\beta_{0}^{*},\beta_{1}^{*})}{\partial\beta_{0}}\right\}  & = & -E\left[\left\{ 1-\frac{1-A}{1-e(X,U;\alpha^{*})}\right\} \frac{\partial\mu_{0}(X;\beta_{0}^{*})}{\partial\beta_{0}}\right],\\
E\left\{ \frac{\partial\hat{\tau}_{\aipw,2}(\alpha^{*},\beta_{0}^{*},\beta_{1}^{*})}{\partial\beta_{1}}\right\}  & = & E\left[\left\{ 1-\frac{A}{e(X,U;\alpha^{*})}\right\} \frac{\partial\mu_{1}(X,U;\beta_{1}^{*})}{\partial\beta_{1}}\right].
\end{eqnarray*}
Under Assumption \ref{asump outcome}, $H_{\aipw}=0$. Under Assumption
\ref{asump ps}, 
\[
\Sigma_{\alpha\alpha}=E\left\{ \frac{\partial S(A,X,U;\alpha^{*})}{\partial\alpha^{\T}}\right\} =E\left\{ S(A,X,U;\alpha^{*})^{\otimes2}\right\} .
\]
Therefore, we can always replace $E\left\{ \partial S(A,X,U;\alpha^{*})/\partial\alpha^{\T}\right\} $
by $E\left\{ S(A,X,U;\alpha^{*})^{\otimes2}\right\} $ in expression
\eqref{eq:line1} if Assumption \ref{asump ps} holds. Thus, we can
derive the influence function for the AIPW estimator.


\section{Lemmas for error-prone estimators}

The error-prone estimators can be viewed as the initial estimators
in Examples \ref{example reg}--\ref{example aipw} with $U$ being
null. The following results are similar to Lemmas \ref{lemma:reg}--\ref{lemma(AIPW)}
with a subtle difference that neither the propensity score or the
outcome model is correctly specified. For completeness, we establish
the results for the error-prone estimators.

Let $\mu_{a}(X;\eta_{a})$ be a working model for $\mu_{a}(X)$, for
$a=0,1$, and $e(X;\gamma)$ be a working model for $e(X)$. Let $S_{a}(A,X,Y;\eta_{a})$
be the estimating function for $\eta_{a}$, e.g., 
\[
S_{a}(A,X,Y;\eta_{a})=\bone(A=a)\frac{\partial\mu_{a}(X;\eta_{a})}{\partial\eta_{a}}\{Y-\mu_{a}(X;\eta_{a})\},
\]
for $a=0,1$. Let $S(A,X;\gamma)$ be the estimating function for
$\gamma$, e.g., 
\[
S(A,X;\gamma)=\frac{A-e(X;\gamma)}{e(X;\gamma)\{1-e(X;\gamma)\}}\frac{\partial e(X;\gamma)}{\partial\gamma}.
\]
We further define 
\[
\Sigma_{\gamma\gamma}=E\left\{ \frac{\partial S(A,X;\gamma^{*})}{\partial\gamma^{\T}}\right\} .
\]
Let $\hat{\eta}_{d,a}$ $(a=0,1)$ and $\hat{\gamma}_{d}$ be the
estimators solving the corresponding estimating equations based on
$\mathcal{S}_{d}$, and let $\eta_{a}^{*}$ $(a=0,1)$ and $\gamma^{*}$
satisfy $E\{S_{a}(A,X,Y;\eta_{a}^{*})\}=0$ $(a=0,1)$ and $E\{S(A,X;\gamma^{*})\}=0$.
Under suitable regularity conditions, $\hat{\eta}_{d,a}$ $(a=0,1)$
and $\hat{\gamma}_{d}$ have probability limits $\eta_{a}^{*}$ $(a=0,1)$
and $\gamma^{*}$.

\begin{lemma}[Regression imputation]\label{Lemma_ep1}The error-prone
regression imputation estimator for $\tau$ is $\hat{\tau}_{\reg,d,\ep}=n_{d}^{-1}\sum_{j\in\mathcal{S}_{d}}\hat{\tau}_{\reg,d,\ep,j},$
where 
\[
\hat{\tau}_{\reg,d,\ep,j}=\mu_{1}(X_{j};\hat{\eta}_{d,1})-\mu_{0}(X_{j};\hat{\eta}_{d,0}).
\]
It has probability limit $\tau_{\ep}=E\left\{ \mu_{1}(X;\eta_{1}^{*})-\mu_{0}(X;\eta_{0}^{*})\right\} $
and influence function 
\begin{eqnarray*}
\phi_{\reg}(A_{j},X_{j},Y_{j}) & = & \mu_{1}(X_{j};\eta_{1}^{*})-\mu_{0}(X_{j};\eta_{0}^{*})-\tau_{\ep}\\
 &  & -E\left\{ \frac{\partial\mu_{1}(X;\eta_{1}^{*})}{\partial\eta_{1}^{\T}}\right\} E\left\{ \frac{\partial S_{1}(A,X,Y;\eta_{1}^{*})}{\partial\eta_{1}^{\T}}\right\} ^{-1}S_{1}(A_{j},X_{j},Y_{j};\eta_{1}^{*})\\
 &  & +E\left\{ \frac{\partial\mu_{0}(X;\eta_{0}^{*})}{\partial\eta_{0}^{\T}}\right\} E\left\{ \frac{\partial S_{0}(A,X,Y;\eta_{0}^{*})}{\partial\eta_{0}^{\T}}\right\} ^{-1}S_{0}(A_{j},X_{j},Y_{j};\eta_{0}^{*}).
\end{eqnarray*}

\end{lemma}

\begin{lemma}[Inverse probability weighting]\label{Lemma_ep2}The
error-prone IPW estimator for $\tau$ is $\hat{\tau}_{\ipw,d,\ep}=n_{d}^{-1}\sum_{j\in\mathcal{S}_{d}}\hat{\tau}_{\ipw,d,\ep,j},$
where 
\[
\hat{\tau}_{\ipw,d,\ep,j}=\frac{A_{j}Y_{j}}{e(X_{j};\hat{\gamma}_{d})}-\frac{(1-A_{j})Y_{j}}{1-e(X_{j};\hat{\gamma}_{d})}.
\]
It has probability limit 
\[
\tau_{\ep}=E\left\{ \frac{AY}{e(X;\gamma^{*})}-\frac{(1-A)Y}{1-e(X;\gamma^{*})}\right\} 
\]
and influence function 
\[
\phi_{\ipw}(A_{j},X_{j},Y_{j})=\left\{ \frac{A_{j}Y_{j}}{e(X_{j};\gamma^{*})}-\frac{(1-A_{j})Y_{j}}{1-e(X_{j};\gamma^{*})}-\tau_{\ep}\right\} -\tilde{H}_{\ipw}\Sigma_{\gamma\gamma}^{-1}S(A_{j},X_{j};\gamma^{*}),
\]
where 
\begin{eqnarray*}
\tilde{H}_{\ipw} & = & E\left(\left[\frac{AY}{e(X;\gamma^{*})^{2}}-\frac{(1-A)Y}{\{1-e(X;\gamma^{*})\}^{2}}\right]\frac{\partial e(X;\gamma^{*})}{\partial\gamma}\right).
\end{eqnarray*}

\end{lemma}

\begin{lemma}[Augmented inverse probability weighting]\label{Lemma_ep3}

Define the residual outcome as $\tilde{R}_{d,j}=Y_{j}-\mu_{1}(X_{j};\hat{\eta}_{d,1})$
for treated units and $\tilde{R}_{d,j}=Y_{j}-\mu_{0}(X_{j};\hat{\eta}_{d,0})$
for control units, for $d=1,2$. The error-prone AIPW estimator for
$\tau$ is $\hat{\tau}_{\aipw,d,\ep}=n_{d}^{-1}\sum_{j\in\mathcal{S}_{d}}\hat{\tau}_{\aipw,d,\ep,j},$
where 
\begin{equation}
\hat{\tau}_{\aipw,d,\ep,j}=\frac{A_{j}\tilde{R}_{d,j}}{e(X_{j};\hat{\gamma}_{d})}+\mu_{1}(X_{j};\hat{\eta}_{d,1})-\frac{(1-A_{j})\tilde{R}_{d,j}}{1-e(X_{j};\hat{\gamma}_{d})}-\mu_{0}(X_{j};\hat{\eta}_{d,0}).\label{eq:aipw ep j}
\end{equation}
It has probability limit 
\[
\tau_{\ep}=E\left[\frac{AY}{e(X;\gamma^{*})}+\left\{ 1-\frac{A}{e(X;\gamma^{*})}\right\} \mu_{1}(X;\eta_{1}^{*})-\frac{(1-A)Y}{1-e(X;\gamma^{*})}-\left\{ 1-\frac{1-A}{1-e(X;\gamma^{*})}\right\} \mu_{0}(X;\eta_{0}^{*})\right]
\]
and influence function 
\begin{eqnarray}
 &  & \phi_{\aipw}(A_{j},X_{j},Y_{j})\nonumber \\
 & = & \frac{A_{j}Y_{j}}{e(X_{j};\gamma^{*})}+\left\{ 1-\frac{A_{j}}{e(X_{j};\gamma^{*})}\right\} \mu_{1}(X_{j};\eta_{1}^{*})\nonumber \\
 &  & -\frac{(1-A_{j})Y_{j}}{1-e(X_{j};\gamma^{*})}-\left\{ 1-\frac{1-A_{j}}{1-e(X_{j};\gamma^{*})}\right\} \mu_{0}(X_{j};\eta_{0}^{*})-\tau_{\ep}-\tilde{H}_{\aipw}\Sigma_{\gamma\gamma}^{-1}S(A_{j},X_{j};\gamma^{*})\nonumber \\
 &  & +E\left[\left\{ 1-\frac{1-A}{1-e(X;\gamma^{*})}\right\} \frac{\partial\mu_{0}(X;\eta_{0}^{*})}{\partial\eta_{0}^{\T}}\right]E\left\{ \frac{\partial S_{0}(A,X,Y;\eta_{0}^{*})}{\partial\eta_{0}^{\T}}\right\} ^{-1}S_{0}(A_{j},X_{j},Y_{j};\eta_{0}^{*})\nonumber \\
 &  & -E\left[\left\{ 1-\frac{A}{e(X;\gamma^{*})}\right\} \frac{\partial\mu_{1}(X;\eta_{1}^{*})}{\partial\eta_{1}^{\T}}\right]E\left\{ \frac{\partial S_{1}(A,X,Y;\eta_{1}^{*})}{\partial\eta_{1}^{\T}}\right\} ^{-1}S_{1}(A_{j},X_{j},Y_{j};\eta_{1}^{*}),\label{eq:ep-aipw}
\end{eqnarray}
where 
\begin{eqnarray*}
\tilde{H}_{\aipw} & = & E\left(\left[\frac{A\{Y-\mu_{1}(X;\eta_{1}^{*})\}}{e(X;\gamma^{*})^{2}}-\frac{(1-A)\{Y-\mu_{0}(X;\eta_{0}^{*})\}}{\{1-e(X;\gamma^{*})\}^{2}}\right]\frac{\partial e(X;\gamma^{*})}{\partial\gamma}\right).
\end{eqnarray*}

\end{lemma}

The results in Lemmas \ref{Lemma_ep1} and \ref{Lemma_ep2} can be
obtained by the Taylor expansion. The proof for Lemma \ref{Lemma_ep3}
is similar to that for Lemma \ref{lemma(AIPW)} and therefore omitted.

\section{Assumptions for the matching estimator}

We review the assumptions for the matching estimators, which can also
be found in \citet{abadie2006large}.

\begin{assumption}[Population distributions]\label{assump1}

(i) $(X,U)$ is continuously distributed on a compact and convex support.
The density of $(X,U)$ is bounded and bounded away from zero on its
support.

(ii) For $a=0,1$, $\mu_{a}(x,u)$ and $\sigma_{a}^{2}(x,u)$ are
Lipschitz, $\sigma_{a}^{2}(x,u)$ is bounded away from zero, and $E(Y^{4}\mid A=a,X=x,U=u)$
is bounded uniformly over its support.

(iii) for $a=0,1$, $\mu_{a}(x)$ and $\sigma_{a}^{2}(x)$ are Lipschitz,
$\sigma_{a}^{2}(x)$ is bounded away from zero, and $E(Y^{4}\mid A=a,X=x)$
is bounded uniformly over its support.

\end{assumption}

Assumption \ref{assump1} (i) can be relaxed by allowing $(X,U)$
to have discrete components. We only need to obtain results on each
level of discrete covariates and derive the same result. Assumption
\ref{assump1} (ii) requires the conditional mean and variance functions
to be bounded and satisfy certain smoothness conditions, which are
rather mild. In fact, Assumption \ref{assump1} (ii) implies Assumption
\ref{assump1} (iii). To be more transparent, we state Assumption
\ref{assump1} (iii) explicitly.

\begin{assumption}[Estimators of mean functions]\label{assump2}For
$a=0,1$, the estimators $\hat{\mu}_{a}(x,u)$ and $\hat{\mu}_{a}(x)$
satisfy the following asymptotic conditions: (i) $|\hat{\mu}_{a}(x,u)-\mu_{a}(x,u)|=o_{P}\left\{ n_{2}^{-1/2+1/\dim(x,u)}\right\} $;
and (ii) $|\hat{\mu}_{a}(x)-\mu_{a}(x)|=o_{P}\left\{ n_{2}^{-1/2+1/\dim(x)}\right\} $.

\end{assumption}

If $\hat{\mu}_{a}(x,u)$ and $\hat{\mu}_{a}(x)$ are obtained under
correctly specified parametric models, then Assumption \ref{assump2}
holds. If $\hat{\mu}_{a}(x,u)$ and $\hat{\mu}_{a}(x)$ are obtained
using nonparametric methods, such as power series regression \citep{newey1997convergence}
or kernel regression \citep{fan1996local} estimators, we need to
select their tuning parameters properly to ensure Assumption \ref{assump2}.
Assumption \ref{assump2} is needed so that the bias correction terms
achieve fast convergence; e.g., $n_{2}^{1/2}(\hat{B}_{2}-B_{2})\rightarrow0$
in probability, as $n_{2}\rightarrow\infty$.

\section{Proof of Theorem \ref{Thm for mat}}

\textcolor{black}{First, we express 
\begin{eqnarray*}
 &  & n_{2}^{1/2}(\hat{\tau}_{\mat,2}-\tau)\\
 & = & n_{2}^{-1/2}\sum_{j\in\mathcal{S}_{2}}\hat{\psi}_{\mat,j}+o_{P}(1)\\
 & = & n_{2}^{-1/2}\sum_{j\in\mathcal{S}_{2}}\left\{ \mu_{1}(X_{j},U_{j})-\mu_{0}(X_{j},U_{j})-\tau\right\} \\
 &  & +n_{2}^{-1/2}\sum_{j\in\mathcal{S}_{2}}(2A_{j}-1)\left\{ 1+M^{-1}K_{2,(X,U),j}\right\} \left\{ Y_{j}-\mu_{A_{j}}(X_{j},U_{j})\right\} +o_{P}(1).
\end{eqnarray*}
Second, let 
\begin{eqnarray*}
T^{\mu} & \equiv & n_{2}^{-1/2}\sum_{j\in\mathcal{S}_{2}}\{\mu_{1}(X_{j},U_{j})-\mu_{0}(X_{j},U_{j})-\tau\},\\
T^{e} & \equiv & n_{2}^{-1/2}\sum_{j\in\mathcal{S}_{2}}(2A_{j}-1)\{1+M^{-1}K_{2,(X,U),j}\}\{Y_{j}-\mu_{A_{j}}(X_{j},U_{j})\},\\
\mathcal{F}_{X,U} & \equiv & \{(X_{j},U_{j}):j\in\mathcal{S}_{2}\}.
\end{eqnarray*}
We verify that the covariance between $T^{\mu}$ and $T^{e}$ is zero:
\begin{eqnarray*}
\cov\left(T^{\mu},T^{e}\right) & = & E\left\{ \cov\left(T^{\mu},T^{e}\mid\mathcal{F}_{X,U}\right)\right\} +\cov\left\{ E(T^{\mu}\mid\mathcal{F}_{X,U}),E(T^{e}\mid\mathcal{F}_{X,U})\right\} \\
 & = & E(0)+\cov\left\{ E(T^{\mu}\mid\mathcal{F}_{X,U}),0\right\} \\
 & = & 0.
\end{eqnarray*}
Ignoring the $o_{P}(1)$ term, the variance of $n_{2}^{1/2}(\hat{\tau}_{\mat,2}-\tau)$
is 
\begin{multline*}
\var\left[n_{2}^{-1/2}\sum_{j\in\mathcal{S}_{2}}\left\{ \mu_{1}(X_{j},U_{j})-\mu_{0}(X_{j},U_{j})-\tau\right\} \right]\\
+\var\left[n_{2}^{-1/2}\sum_{j\in\mathcal{S}_{2}}(2A_{j}-1)\left\{ 1+M^{-1}K_{2,(X,U),j}\right\} \left\{ Y_{j}-\mu_{A_{j}}(X_{j},U_{j})\right\} \right].
\end{multline*}
As $n_{2}\rightarrow\infty$, the first term becomes 
\[
v_{2}^{\tau}=\var\left\{ \tau(X,U)\right\} ,
\]
and the second term becomes 
\[
v_{2}^{e}=\plim\left[n_{2}^{-1}\sum_{j\in\mathcal{S}_{2}}\left\{ 1+M^{-1}K_{2,(X,U),j}\right\} ^{2}\sigma_{A_{j}}^{2}(X_{j},U_{j})\right].
\]
Under Assumption \ref{assump1}, $K_{2,(X,U),j}=O_{P}(1)$ and $E\{K_{2,(X,U),j}\}$
and $E\{K_{2,(X,U),j}^{2}\}$ are uniformly bounded over $n_{2}$
(\citealp{abadie2006large}, Lemma 3). Therefore, a simple algebra
yields $v_{2}^{e}=O(1)$. Combining all results, the asymptotic variance
of $n_{2}^{1/2}(\hat{\tau}_{\mat,2}-\tau)$ is $v_{2}^{\tau}+v_{2}^{e}$. }

The derivations for $\Gamma$ and $V$ are similar and thus omitted.

\section{Proof of Theorem \ref{Thm: boot} \label{sec:Proof-of-Boot}}

For the matching estimators, \citet{otsu2016bootstrap} showed that
the distribution of $\hat{\tau}_{2}^{*}-\hat{\tau}_{2}$ given the
observed data approximates the sampling distribution of $\hat{\tau}_{2}$.
In what follows, we prove that $\widehat{\var}(\hat{\tau}_{2})$ is
consistent for $\var(\hat{\tau}_{2})$, which covers both cases with
RAL and matching estimators and is simpler than \citet{otsu2016bootstrap}
for the matching estimators.

Let $(M_{1},\ldots,M_{n_{2}})$ be a multinomial random vector with
$n_{2}$ draws on $n_{2}$ cells with equal probabilities. Let $W_{j}^{*}=n_{2}^{-1/2}M_{j}$
for $j=1,\ldots,n_{2}$, and $\bar{W}^{*}=n_{2}^{-1}\sum_{j\in\mathcal{S}_{2}}W_{j}^{*}$.
Then, the bootstrap weights $\{W_{j}^{*}:j=1,\ldots,n_{2}\}$ satisfy
that as $n_{2}\rightarrow\infty$, 
\begin{equation}
\max_{j\in\mathcal{S}_{2}}|W_{j}^{*}-\bar{W}^{*}|\stackrel{P}{\rightarrow}0,\label{eq:bootweight1}
\end{equation}
\begin{equation}
\sum_{j\in\mathcal{S}_{2}}(W_{j}^{*}-\bar{W}^{*})^{2}\stackrel{P}{\rightarrow}1.\label{eq:bootweight2}
\end{equation}
See, e.g., \citet{mason1992rank}. The bootstrap replicate of $(\hat{\tau}_{2}-\tau)$
can be written as 
\begin{equation}
\hat{\tau}_{2}^{*}-\hat{\tau}_{2}=n_{2}^{-1/2}\sum_{j\in\mathcal{S}_{2}}\left(W_{j}^{*}-\bar{W}^{*}\right)\hat{\psi}_{j}.\label{eq:boot1}
\end{equation}

For RAL estimators, following (\ref{eq:boot1}), we have $\hat{\tau}_{2}^{*}-\hat{\tau}_{2}=T_{1}^{*}+T_{2}^{*},$
where 
\begin{eqnarray*}
T_{1}^{*} & = & n_{2}^{-1/2}\sum_{j\in\mathcal{S}_{2}}\left(W_{j}^{*}-\bar{W}^{*}\right)\psi(A_{j},X_{j},U_{j},Y_{j}),\\
T_{2}^{*} & = & n_{2}^{-1/2}\sum_{j\in\mathcal{S}_{2}}\left(W_{j}^{*}-\bar{W}^{*}\right)\left\{ \hat{\psi}(A_{j},X_{j},U_{j},Y_{j})-\psi(A_{j},X_{j},U_{j},Y_{j})\right\} .
\end{eqnarray*}
By (\ref{eq:bootweight1}) and the fact that $\hat{\psi}(A_{j},X_{j},U_{j},Y_{j})$
is root-$n$ consistent for $\psi(A_{j},X_{j},U_{j},Y_{j})$, we have
$T_{2}^{*}=o_{P}\left(n_{2}^{-1}\right)$. By (\ref{eq:bootweight2}),
we have 
\begin{equation}
\hat{\tau}_{2}^{*}-\hat{\tau}_{2}=T_{1}^{*}+o_{P}(1)=n_{2}^{-1/2}\sum_{j\in\mathcal{S}_{2}}\left(W_{j}^{*}-\bar{W}^{*}\right)\psi(A_{j},X_{j},U_{j},Y_{j})+o_{P}(1).\label{eq:boot2}
\end{equation}

For matching estimators, following (\ref{eq:boot1}), we have

\begin{eqnarray*}
\hat{\tau}_{2}^{*}-\hat{\tau}_{2} & = & n_{2}^{-1/2}\sum_{j\in\mathcal{S}_{2}}\left(W_{j}^{*}-\bar{W}^{*}\right)\hat{\psi}_{\mat,j}\\
 & \equiv & T_{\mat,1}^{*}+T_{\mat,2}^{*}+T_{\mat,3}^{*},
\end{eqnarray*}
where 
\begin{eqnarray*}
T_{\mat,1}^{*} & = & n_{2}^{-1/2}\sum_{j\in\mathcal{S}_{2}}\left(W_{j}^{*}-\bar{W}^{*}\right)\psi_{\mat,j},\\
T_{\mat,2}^{*} & = & n_{2}^{-1/2}\sum_{j\in\mathcal{S}_{2}}\left(W_{j}^{*}-\bar{W}^{*}\right)\left[\left\{ \hat{\mu}_{1}(X_{j},U_{j})-\hat{\mu}_{0}(X_{j},U_{j})-\hat{\tau}_{2}\right\} -\left\{ \mu_{1}(X_{j},U_{j})-\mu_{0}(X_{j},U_{j})-\tau\right\} \right],\\
T_{\mat,3}^{*} & = & n_{2}^{-1/2}\sum_{j\in\mathcal{S}_{2}}\left(W_{j}^{*}-\bar{W}^{*}\right)(2A_{j}-1)\left\{ 1+M^{-1}K_{2,(X,U),j}\right\} \left\{ \mu_{A_{j}}(X_{j},U_{j})-\hat{\mu}_{A_{j}}(X_{j},U_{j})\right\} .
\end{eqnarray*}
Under Assumption \ref{assump1}, for any $\zeta>0$, $E\{K_{2,(X,U),j}^{\zeta}\}$
is uniformly bounded over $n_{2}$ \citep{abadie2006large}. Together
with Assumption \ref{assump2} and the property of the bootstrap weights
that $\max_{j\in\mathcal{S}_{2}}|W_{j}^{*}-\bar{W}^{*}|\rightarrow0$
in probability, as $n_{2}\rightarrow\infty$, we have $T_{\mat,2}^{*}=o_{P}\left\{ n_{2}^{-1+1/\dim(x,u)}\right\} $
and $T_{\mat,3}^{*}=o_{P}\left\{ n_{2}^{-1+1/\dim(x,u)}\right\} $.
By (\ref{eq:bootweight2}), we have 
\begin{equation}
\hat{\tau}_{2}^{*}-\hat{\tau}_{2}=T_{\mat,1}^{*}+o_{P}(1)=n_{2}^{-1/2}\sum_{j\in\mathcal{S}_{2}}\left(W_{j}^{*}-\bar{W}^{*}\right)\psi_{\mat,j}+o_{P}(1).\label{eq:boot3}
\end{equation}

Unifying (\ref{eq:boot2}) and (\ref{eq:boot3}), $\hat{\tau}_{2}^{*}-\hat{\tau}_{2}=n_{2}^{-1/2}\sum_{j\in\mathcal{S}_{2}}\left(W_{j}^{*}-\bar{W}^{*}\right)\psi_{j}+o_{P}(1).$
Let $\psi_{(i)}$ be the $i$th order statistic of \textcolor{black}{$\{\psi_{j}:j\in\mathcal{S}_{2}\}$.
Because $E\left(\psi_{j}^{\gamma}\right)<\infty$} for $0\leq\gamma\leq4$,
we have 
\[
\frac{|\hat{\tau}_{2}^{*}-\hat{\tau}_{2}|}{n_{2}^{1/\gamma}}\leq2\frac{|\psi_{(1)}|+|\psi_{(n_{2})}|}{n_{2}^{1/\gamma}}\rightarrow0,
\]
almost surely, as $n_{2}\rightarrow\infty$, leading to $\max_{\{W_{j}^{*}:j\in\mathcal{S}_{2}\}}|\hat{\tau}_{2}^{*}-\hat{\tau}_{2}|/n_{2}^{1/\gamma}\rightarrow0$,
almost surely, as $n_{2}\rightarrow\infty$, where the maximum is
taken over all possible bootstrap replicates. By Theorem 3.8 of \citet{shao2012jackknife},
\[
\frac{\widehat{\var}(\hat{\tau}_{2})}{\var(\hat{\tau}_{2})}\rightarrow1,
\]
almost surely, as $n_{2}\rightarrow\infty$. This proves that $\widehat{\var}(\hat{\tau}_{2})$
is consistent for $\var(\hat{\tau}_{2})$.

The proofs for the consistency of $\hat{\Gamma}$ and $\hat{V}$ for
$\Gamma$ and $V$ are similar and thus omitted. Therefore, $\widehat{\var}(\hat{\tau})$
is consistent for $\var(\hat{\tau})$.

\section{Derivation of the optimal sample allocation }

The goal is to minimize (\ref{eq:var}) subject to the constraint
(\ref{eq:budget}). By the Lagrange multipliers technique, it suffices
to find the minimizer of 
\[
\mathcal{L}(n_{1},n_{2},\lambda)=n_{2}^{-1}v_{2}-(n_{2}^{-1}-n_{1}^{-1})\gamma-\lambda(C-n_{1}C_{1}-n_{2}C_{2}).
\]
Solving $\partial\mathcal{L}(n_{1},n_{2},\lambda)/\partial n_{1}=0$
and $\partial\mathcal{L}(n_{1},n_{2},\lambda)/\partial n_{2}=0$,
we have 
\[
n_{1}=\left(\frac{\gamma}{\lambda C_{1}}\right)^{1/2},\quad\quad n_{2}=\left(\frac{v_{2}-\gamma}{\lambda C_{2}}\right)^{1/2},
\]
and therefore, (\ref{eq:optimal sample alloc}) follows.

\section{Proof of Theorem \ref{Thm:relaxRAL}}

Following the discussion in Section \ref{subsec:RelaxationRAL}, we
can express 
\begin{eqnarray*}
n_{2}^{1/2}(\hat{\tau}_{2}-\tau) & = & \rho^{1/2}n_{1}^{-1/2}\sum_{j\in\mathcal{S}_{1}}\pi_{j}^{-1}I_{j}\psi(A_{j},X_{j},U_{j},Y_{j})+o_{P}(1),\\
n_{2}^{1/2}(\hat{\tau}_{2,\ep}-\hat{\tau}_{1,\ep}) & = & \rho^{1/2}n_{1}^{-1/2}\sum_{j\in\mathcal{S}_{1}}(\pi_{j}^{-1}I_{j}-1)\phi(A_{j},X_{j},Y_{j})+o_{P}(1).
\end{eqnarray*}
Then, the asymptotic normality of $\{n_{2}^{1/2}(\hat{\tau}_{2}-\tau),n_{2}^{1/2}(\hat{\tau}_{2,\ep}-\hat{\tau}_{1,\ep})\}^{\T}$
follows from the multivariate central limit theorem. The corresponding
asymptotic covariance matrix can be obtained by the following calculation.
First, $v_{2}$ is 
\begin{eqnarray*}
 &  & \var\left\{ \rho^{1/2}n_{1}^{-1/2}\sum_{j\in\mathcal{S}_{1}}\pi_{j}^{-1}I_{j}\psi(A_{j},X_{j},U_{j},Y_{j})\right\} \\
 & = & \rho\times\var\{\pi^{-1}I\psi(A,X,U,Y)\}.
\end{eqnarray*}
Second, $\Gamma$ is 
\begin{eqnarray*}
 &  & \cov\left\{ \rho^{1/2}n_{1}^{-1/2}\sum_{j\in\mathcal{S}_{1}}\pi_{j}^{-1}I_{j}\psi(A_{j},X_{j},U_{j},Y_{j}),\rho^{1/2}n_{1}^{-1/2}\sum_{j\in\mathcal{S}_{1}}(\pi_{j}^{-1}I_{j}-1)\phi(A_{j},X_{j},Y_{j})\right\} \\
 & = & \rho\times\cov\{\pi^{-1}I\psi(A,X,U,Y),(\pi^{-1}I-1)\phi(A,X,Y)\}.
\end{eqnarray*}
Third, $V$ is 
\begin{eqnarray*}
 &  & \var\left\{ \rho^{1/2}n_{1}^{-1/2}\sum_{j\in\mathcal{S}_{1}}(\pi_{j}^{-1}I_{j}-1)\phi(A_{j},X_{j},Y_{j})\right\} \\
 & = & \rho\times\var\{(\pi^{-1}I-1)\phi(A,X,Y)\}.
\end{eqnarray*}

\section{Error-prone matching estimators in Section \ref{subsec:Relaxation-mat}}

\textcolor{black}{Let the error-prone matching estimators be 
\begin{eqnarray*}
\hat{\tau}_{\mat,1,\ep}^{(0)} & = & n_{1}^{-1}\sum_{i\in\mathcal{S}_{1}}(2A_{i}-1)\left(Y_{i}-\frac{1}{M}\sum_{l\in\J_{1,X,i}}Y_{l}\right),\\
\hat{\tau}_{\mat,2,\ep}^{(0)} & = & \left(\sum_{j\in\mathcal{S}_{2}}\pi_{j}^{-1}\right)^{-1}\sum_{j\in\mathcal{S}_{2}}\pi_{j}^{-1}(2A_{j}-1)\left(Y_{j}-\frac{1}{M}\sum_{l\in\J_{2,X,j}}Y_{l}\right).
\end{eqnarray*}
Let the bias-corrected matching estimators be 
\[
\hat{\tau}_{\mat,d,\ep}=\hat{\tau}_{\mat,d,\ep}^{(0)}-n_{1}^{-1/2}\hat{B}_{d,\ep},\quad(d=1,2),
\]
where 
\[
\hat{B}_{d,\ep}=n_{1}^{-1/2}\sum_{j\in\mathcal{S}_{2}}\pi_{j}^{-1}(2A_{j}-1)\left[\frac{1}{M}\sum_{l\in\J_{d,X,j}}\left\{ \hat{\mu}_{1-A_{j}}(X_{j})-\hat{\mu}_{1-A_{j}}(X_{l})\right\} \right].
\]
}

\section{Proof of Theorem \ref{Thm:relaxMat}}

First, with the new definition (\ref{eq:mat2 new}), we can expres\textcolor{black}{s
\begin{eqnarray*}
 &  & n_{2}^{1/2}(\hat{\tau}_{\mat,2}-\tau)\\
 & = & n_{2}^{1/2}(\hat{\tau}_{\mat,2}^{(0)}-n_{1}^{-1/2}\hat{B}_{2}-\tau)\\
 & = & n_{2}^{1/2}\left\{ \left(\sum_{j\in\mathcal{S}_{2}}\pi_{j}^{-1}\right)^{-1}\sum_{j\in\mathcal{S}_{2}}\pi_{j}^{-1}(2A_{j}-1)\left\{ 1+M^{-1}K_{2,(X,U),j}\right\} Y_{j}-\tau\right\} \\
 &  & -n_{2}^{1/2}n_{1}^{-1}\sum_{j\in\mathcal{S}_{2}}\pi_{j}^{-1}(2A_{j}-1)\left[M^{-1}\sum_{l\in\J_{2,(X,U),j}}\left\{ \hat{\mu}_{1-A_{j}}(X_{j},U_{j})-\hat{\mu}_{1-A_{j}}(X_{l},U_{l})\right\} \right]\\
 & = & \rho^{1/2}n_{1}^{-1/2}\sum_{j\in\mathcal{S}_{2}}\pi_{j}^{-1}\left\{ \mu_{1}(X_{j},U_{j})-\mu_{0}(X_{j},U_{j})-\tau\right\} \\
 &  & +\rho^{1/2}n_{1}^{-1/2}\sum_{j\in\mathcal{S}_{2}}\pi_{j}^{-1}(2A_{j}-1)\left\{ 1+M^{-1}K_{2,(X,U),j}\right\} \left\{ Y_{j}-\mu_{A_{j}}(X_{j},U_{j})\right\} +o_{P}(1),
\end{eqnarray*}
} where the third equality follows by some algebra.

\textcolor{black}{Second, let 
\begin{eqnarray*}
T^{\mu} & \equiv & n_{1}^{-1/2}\sum_{j\in\mathcal{S}_{2}}\pi_{j}^{-1}\{\mu_{1}(X_{j},U_{j})-\mu_{0}(X_{j},U_{j})-\tau\},\\
T^{e} & \equiv & n_{1}^{-1/2}\sum_{j\in\mathcal{S}_{2}}\pi_{j}^{-1}(2A_{j}-1)\{1+M^{-1}K_{2,(X,U),j}\}\{Y_{j}-\mu_{A_{j}}(X_{j},U_{j})\},\\
\mathcal{F}_{X,U} & \equiv & \{(X_{j},U_{j}):j\in\mathcal{S}_{2}\}.
\end{eqnarray*}
We verify that the covariance between $T^{\mu}$ and $T^{e}$ is zero:
\begin{eqnarray*}
\cov\left(T^{\mu},T^{e}\right) & = & E\left\{ \cov\left(T^{\mu},T^{e}\mid\mathcal{F}_{X,U}\right)\right\} +\cov\left\{ E(T^{\mu}\mid\mathcal{F}_{X,U}),E(T^{e}\mid\mathcal{F}_{X,U})\right\} \\
 & = & E(0)+\cov\left\{ E(T^{\mu}\mid\mathcal{F}_{X,U}),0\right\} \\
 & = & 0.
\end{eqnarray*}
Ignoring the $o_{P}(1)$ term, the variance of $n_{2}^{1/2}(\hat{\tau}_{\mat,2}-\tau)$
is 
\begin{multline*}
\rho\times\var\left[n_{1}^{-1/2}\sum_{j\in\mathcal{S}_{2}}\pi_{j}^{-1}\left\{ \mu_{1}(X_{j},U_{j})-\mu_{0}(X_{j},U_{j})-\tau\right\} \right]\\
+\rho\times\var\left[n_{1}^{-1/2}\sum_{j\in\mathcal{S}_{2}}\pi_{j}^{-1}(2A_{j}-1)\left\{ 1+M^{-1}K_{2,(X,U),j}\right\} \left\{ Y_{j}-\mu_{A_{j}}(X_{j},U_{j})\right\} \right].
\end{multline*}
As $n_{2}\rightarrow\infty$, the first term becomes 
\[
\tilde{v}_{2}^{\tau}=\rho\times E\left[\frac{1-\pi}{\pi}\left\{ \tau(X,U)-\tau\right\} ^{2}\right],
\]
and the second term becomes 
\[
\tilde{v}_{2}^{e}=\rho\times\plim\left[n_{1}^{-1}\sum_{j\in\mathcal{S}_{1}}\frac{1-\pi_{j}}{\pi_{j}}\left\{ 1+M^{-1}K_{2,(X,U),j}\right\} ^{2}\sigma_{A_{j}}^{2}(X_{j},U_{j})\right].
\]
Because $\pi_{j}$'s are bounded away from zero, under Assumption
\ref{assump1}, we have $K_{2,(X,U),j}=O_{P}(1)$, and $E\{K_{2,(X,U),j}\}$
and $E\{K_{2,(X,U),j}^{2}\}$ are uniformly bounded over $n_{2}$
(\citealp{abadie2006large}, Lemma 3). Therefore, a simple algebra
yields $\tilde{v}_{2}^{e}=O(1)$. Combining all results, the asymptotic
variance of $n_{2}^{1/2}(\hat{\tau}_{\mat,2}-\tau)$ is $\tilde{v}_{2}^{\tau}+\tilde{v}_{2}^{e}$. }

The derivations for $\Gamma$ and $V$ are similar and thus omitted.

\section{Proof of (\ref{eq:IF_for_RAL2})}

We write $\hat{\tau}_{2}=\hat{\tau}_{2}(\hat{\alpha},\hat{\beta}_{0},\hat{\beta}_{1})$
and $\hat{\tau}_{\aipw,2,j}=\hat{\tau}_{\aipw,2,j}(\hat{\alpha},\hat{\beta}_{0},\hat{\beta}_{1})$
to emphasize its dependence on the parameter estimates $(\hat{\alpha},\hat{\beta}_{0},\hat{\beta}_{1})$.

First, we write 
\begin{eqnarray*}
\hat{\tau}_{2}(\alpha^{*},\beta_{0}^{*},\beta_{1}^{*}) & = & \frac{\sum_{j\in\mathcal{S}_{2}}\pi_{j}^{-1}\hat{\tau}_{\aipw,2,j}(\alpha^{*},\beta_{0}^{*},\beta_{1}^{*})}{\sum_{j\in\mathcal{S}_{2}}\pi_{j}^{-1}}\\
 & = & \frac{n_{1}^{-1}\sum_{j\in\mathcal{S}_{1}}\pi_{j}^{-1}I_{j}\hat{\tau}_{\aipw,2,j}(\alpha^{*},\beta_{0}^{*},\beta_{1}^{*})}{n_{1}^{-1}\sum_{j\in\mathcal{S}_{1}}\pi_{j}^{-1}I_{j}}\equiv\frac{\hat{T}_{1}}{\hat{T}_{2}},
\end{eqnarray*}
where $\hat{T}_{1}\equiv n_{1}^{-1}\sum_{j\in\mathcal{S}_{1}}\pi_{j}^{-1}I_{j}\hat{\tau}_{\aipw,2,j}(\alpha^{*},\beta_{0}^{*},\beta_{1}^{*})$
and $\hat{T}_{2}\equiv n_{1}^{-1}\sum_{j\in\mathcal{S}_{1}}\pi_{j}^{-1}I_{j}$.
Then, $\hat{T}_{1}$ is consistent for $T_{1}\equiv\tau$, and $\hat{T}_{2}$
is consistent for $T_{2}\equiv1$. By the Taylor expansion, we have
\begin{eqnarray*}
\hat{\tau}_{2}(\alpha^{*},\beta_{0}^{*},\beta_{1}^{*})-\tau & \cong & \frac{T_{1}}{T_{2}}+\frac{1}{T_{2}}(\hat{T}_{1}-T_{1})-\frac{T_{1}}{T_{2}^{2}}(\hat{T}_{2}-T_{2})-\tau,\\
 & = & \left\{ n_{1}^{-1}\sum_{j\in\mathcal{S}_{1}}\pi_{j}^{-1}I_{j}\hat{\tau}_{\aipw,2,j}(\alpha^{*},\beta_{0}^{*},\beta_{1}^{*})-\tau\right\} -\tau\left(n_{1}^{-1}\sum_{j\in\mathcal{S}_{1}}\pi_{j}^{-1}I_{j}-1\right)\\
 & = & n_{1}^{-1}\sum_{j\in\mathcal{S}_{1}}\pi_{j}^{-1}I_{j}\left\{ \hat{\tau}_{\aipw,2,j}(\alpha^{*},\beta_{0}^{*},\beta_{1}^{*})-\tau\right\} .
\end{eqnarray*}

Second, by the Taylor expansion, 
\begin{eqnarray*}
0 & = & n_{1}^{-1}\sum_{j\in\mathcal{S}_{2}}\pi_{j}^{-1}S(A_{j},X_{j},U_{j};\hat{\alpha})\\
 & \cong & n_{1}^{-1}\sum_{j\in\mathcal{S}_{2}}\pi_{j}^{-1}S(A_{j},X_{j},U_{j};\alpha^{*})+\left\{ n_{1}^{-1}\sum_{j\in\mathcal{S}_{2}}\pi_{j}^{-1}\frac{\partial S(A_{j},X_{j},U_{j};\alpha^{*})}{\partial\alpha^{\T}}\right\} (\hat{\alpha}-\alpha^{*})\\
 & \cong & n_{1}^{-1}\sum_{j\in\mathcal{S}_{2}}\pi_{j}^{-1}S(A_{j},X_{j},U_{j};\alpha^{*})+E\left\{ \frac{\partial S(A,X,U;\alpha^{*})}{\partial\alpha^{\T}}\right\} (\hat{\alpha}-\alpha^{*}),
\end{eqnarray*}
where the last line follows because by Assumption \ref{asump-missing},
$n_{1}^{-1}\sum_{j\in\mathcal{S}_{2}}\pi_{j}^{-1}\partial S(A_{j},X_{j},U_{j};\alpha^{*})/\partial\alpha^{\T}$
converges to $E\left\{ \partial S(A,X,U;\alpha^{*})/\partial\alpha^{\T}\right\} $
in probability. Therefore, we can establish 
\begin{eqnarray}
\hat{\alpha}-\alpha^{*} & \cong & n_{1}^{-1}\sum_{j\in\mathcal{S}_{2}}E\left\{ \frac{\partial S(A,X,U;\alpha^{*})}{\partial\alpha^{\T}}\right\} ^{-1}\pi_{j}^{-1}S(A_{j},X_{j},U_{j};\alpha^{*}).\label{eq:alpha-hat}
\end{eqnarray}
Similarly, we can establish 
\begin{equation}
\hat{\beta}_{a}-\beta_{a}^{*}\cong n_{1}^{-1}\sum_{j\in\mathcal{S}_{2}}E\left\{ \frac{\partial S_{a}(A,X,U,Y;\beta_{a}^{*})}{\partial\beta_{a}^{\T}}\right\} ^{-1}\pi_{j}^{-1}S_{a}(A_{j},X_{j},U_{j},Y_{j};\beta_{a}^{*})\quad(a=0,1).\label{eq:beta-hat}
\end{equation}

Third, by the Taylor expansion and (\ref{eq:alpha-hat}) and (\ref{eq:beta-hat}),
we obtain 
\begin{eqnarray}
 &  & \hat{\tau}_{2}(\hat{\alpha},\hat{\beta}_{0},\hat{\beta}_{1})\nonumber \\
 & \cong & \hat{\tau}_{2}(\alpha^{*},\beta_{0}^{*},\beta_{1}^{*})+E\left\{ \frac{\partial\hat{\tau}_{2}(\alpha^{*},\beta_{0}^{*},\beta_{1}^{*})}{\partial\alpha^{\T}}\right\} (\hat{\alpha}-\alpha^{*})\nonumber \\
 &  & +E\left\{ \frac{\partial\hat{\tau}_{2}(\alpha^{*},\beta_{0}^{*},\beta_{1}^{*})}{\partial\beta_{0}^{\T}}\right\} (\hat{\beta}_{0}-\beta_{0}^{*})+E\left\{ \frac{\partial\hat{\tau}_{2}(\alpha^{*},\beta_{0}^{*},\beta_{1}^{*})}{\partial\beta_{1}^{\T}}\right\} (\hat{\beta}_{1}-\beta_{1}^{*})\nonumber \\
 & \cong & \hat{\tau}_{2}(\alpha^{*},\beta_{0}^{*},\beta_{1}^{*})\nonumber \\
 &  & +n_{1}^{-1}\sum_{j\in\mathcal{S}_{1}}E\left\{ \frac{\partial\hat{\tau}_{2}(\alpha^{*},\beta_{0}^{*},\beta_{1}^{*})}{\partial\alpha^{\T}}\right\} E\left\{ \frac{\partial S(A,X,U;\alpha^{*})}{\partial\alpha^{\T}}\right\} ^{-1}\pi_{j}^{-1}I_{j}S(A_{j},X_{j},U_{j};\alpha^{*})\label{eq:line2-1}\\
 &  & -n_{1}^{-1}\sum_{j\in\mathcal{S}_{1}}E\left\{ \frac{\partial\hat{\tau}_{2}(\alpha^{*},\beta_{0}^{*},\beta_{1}^{*})}{\partial\beta_{0}^{\T}}\right\} E\left\{ \frac{\partial S_{0}(A,X,U,Y;\beta_{0}^{*})}{\partial\beta_{0}^{\T}}\right\} ^{-1}\pi_{j}^{-1}I_{j}S_{0}(A_{j},X_{j},U_{j},Y_{j};\beta_{0}^{*})\nonumber \\
 &  & -n_{1}^{-1}\sum_{j\in\mathcal{S}_{1}}E\left\{ \frac{\partial\hat{\tau}_{2}(\alpha^{*},\beta_{0}^{*},\beta_{1}^{*})}{\partial\beta_{1}^{\T}}\right\} E\left\{ \frac{\partial S_{1}(A,X,U,Y;\beta_{0}^{*})}{\partial\beta_{1}^{\T}}\right\} ^{-1}\pi_{j}^{-1}I_{j}S_{1}(A_{j},X_{j},U_{j},Y_{j};\beta_{1}^{*}).\nonumber 
\end{eqnarray}
Moreover, we have the following calculations: 
\begin{eqnarray*}
E\left\{ \frac{\partial\hat{\tau}_{2}(\alpha^{*},\beta_{0}^{*},\beta_{1}^{*})}{\partial\alpha}\right\}  & = & -H_{\aipw},\\
E\left\{ \frac{\partial\hat{\tau}_{2}(\alpha^{*},\beta_{0}^{*},\beta_{1}^{*})}{\partial\beta_{0}}\right\}  & = & -E\left[\left\{ 1-\frac{1-A}{1-e(X,U;\alpha^{*})}\right\} \frac{\partial\mu_{0}(X;\beta_{0}^{*})}{\partial\beta_{0}}\right],\\
E\left\{ \frac{\partial\hat{\tau}_{2}(\alpha^{*},\beta_{0}^{*},\beta_{1}^{*})}{\partial\beta_{1}}\right\}  & = & E\left[\left\{ 1-\frac{A}{e(X,U;\alpha^{*})}\right\} \frac{\partial\mu_{1}(X,U;\beta_{1}^{*})}{\partial\beta_{1}}\right].
\end{eqnarray*}
Under Assumption \ref{asump ps}, 
\[
E\left\{ \frac{\partial S(A,X,U;\alpha^{*})}{\partial\alpha^{\T}}\right\} =E\left\{ S(A,X,U;\alpha^{*})^{\otimes2}\right\} =\Sigma_{\alpha\alpha}.
\]
Under Assumption \ref{asump outcome}, we can still replace $E\left\{ \partial S(A,X,U;\alpha^{*})/\partial\alpha^{\T}\right\} $
by $\Sigma_{\alpha\alpha}$, because $H_{\aipw}=0$ and the term in
(\ref{eq:line2-1}) is zero. Therefore, combining the above results,
we have $\hat{\tau}_{2}-\tau=\hat{\tau}_{2}(\hat{\alpha},\hat{\beta}_{0},\hat{\beta}_{1})-\tau=n_{1}^{-1}\sum_{j\in\mathcal{S}_{1}}\pi_{j}^{-1}I_{j}\psi(A_{j},X_{j},U_{j},Y_{j})$\textcolor{black}{,}
where $\psi(A,X,U,Y)$ is given by (\ref{eq:aipw influence fctn}).
The result follows.

\section{Proof of Proposition \ref{prop1}}

According to (\ref{eq:IF_for_RAL2}) and (\ref{eq:IF_for_RALep})
for RAL estimators or (\ref{eq:pi-matching}) and (\ref{eq:pi-matching ep})
for the matching estimators, we express 
\begin{eqnarray*}
\hat{\tau}-\tau & = & \hat{\tau}_{2}-\tau-\hat{\Gamma}^{\T}\hat{V}^{-1}(\hat{\tau}_{2,\ep}-\hat{\tau}_{1,\ep})\\
 & \cong & n_{1}^{-1}\sum_{i\in\mathcal{S}_{1}}\pi_{i}^{-1}I_{i}\psi_{i}-\Gamma^{\T}V^{-1}\left(n_{1}^{-1}\sum_{i\in\mathcal{S}_{1}}\pi_{i}^{-1}I_{i}\phi_{i}-n_{1}^{-1}\sum_{i\in\mathcal{S}_{1}}\phi_{i}\right)\\
 & \cong & n_{1}^{-1}\sum_{i\in\mathcal{S}_{1}}\left\{ \frac{I_{i}}{\pi_{i}}\psi_{i}-\left(\frac{I_{i}}{\pi_{i}}-1\right)\Gamma V^{-1}\phi_{i}\right\} .
\end{eqnarray*}

\section{Proof of Theorem \ref{Thm: boot-1}}

Let $(M_{1},\ldots,M_{n_{1}})$ be a multinomial random vector with
$n_{1}$ draws on $n_{1}$ cells with equal probabilities. Let $W_{i}^{*}=n_{1}^{-1/2}M_{i}$
for $i=1,\ldots,n_{1}$, and $\bar{W}^{*}=n_{1}^{-1}\sum_{i\in\mathcal{S}_{1}}W_{i}^{*}$.
Then, the bootstrap replicate of $(\hat{\tau}_{2}-\tau)$ can be written
as

\begin{eqnarray*}
\hat{\tau}_{2}^{*}-\hat{\tau} & = & n_{1}^{-1/2}\sum_{i\in\mathcal{S}_{1}}\left(W_{i}^{*}-\bar{W}^{*}\right)\pi_{i}^{-1}I_{i}\hat{\psi}_{i}\\
 & \cong & n_{1}^{-1/2}\sum_{i\in\mathcal{S}_{1}}\left(W_{i}^{*}-\bar{W}^{*}\right)\pi_{i}^{-1}I_{i}\psi_{i},
\end{eqnarray*}
where the last line follows by a similar argument as in Section \ref{sec:Proof-of-Boot}
for both RAL and matching estimators.

Let $l_{i}$ denote $\pi_{i}^{-1}I_{i}\psi_{i}$ and $l_{(i)}$ be
the $i$th order statistic of \textcolor{black}{$\{l_{i}:i\in\mathcal{S}_{1}\}$.
Because $E\left(l_{j}^{\gamma}\right)<\infty$} for $0\leq\gamma\leq4$,
we have 
\[
\frac{|\hat{\tau}_{2}^{*}-\hat{\tau}|}{n_{1}^{1/\gamma}}\leq2\frac{|l_{(1)}|+|l_{(n_{1})}|}{n_{1}^{1/\gamma}}\rightarrow0,
\]
almost surely, as $n_{1}\rightarrow\infty$, leading to $\max_{\{W_{i}^{*}:i\in\mathcal{S}_{1}\}}|\hat{\tau}_{2}^{*}-\hat{\tau}|/n_{1}^{1/\gamma}\rightarrow0$,
almost surely, as $n_{1}\rightarrow\infty$, where the maximum is
taken over all possible bootstrap replicates. By Theorem 3.8 of \citet{shao2012jackknife},
\[
\frac{\widehat{\var}(\hat{\tau}_{2})}{\var(\hat{\tau}_{2})}\rightarrow1,
\]
almost surely, as $n_{1}\rightarrow\infty$. This proves that $\widehat{\var}(\hat{\tau}_{2})$
is consistent for $\var(\hat{\tau}_{2})$.

The proofs for the consistency of $\hat{\Gamma}$ and $\hat{V}$ for
$\Gamma$ and $V$ are similar and thus omitted. Therefore, $\widehat{\var}(\hat{\tau})$
is consistent for $\var(\hat{\tau})$.
\end{document}